\documentclass[aps,twocolumn,showpacs,amsmath,amssymb,superscriptaddress,prc]{revtex4-2}
\usepackage{graphicx,here}
\usepackage{multirow}
\usepackage{tikz}
\usepackage{epstopdf}
\usepackage[percent]{overpic}
\usepackage{scrextend}
\usepackage{xcolor,xspace}
\usepackage[ulem=normalem]{changes}
\usepackage{hyperref}
\hypersetup{colorlinks=True,urlcolor=blue,linkcolor=blue,citecolor=blue,filecolor=black}

\colorlet{Changes@Color}{red}
\usepackage{dcolumn,color,footnote,bm,braket}
\usepackage{url,longtable,tabularx}
\usepackage{threeparttable}

\usepackage{fancybox}
\usetikzlibrary{matrix}

\newcommand\+{\dagger}
\newcommand\jr{j_{\rho}}
\newcommand\mr{m_{\rho}}

\newcommand\mgtbeta{M({\mathrm{GT}})}
\newcommand\mfbeta{M({\mathrm{F}})}

\newcommand\hb{\hat{H}_{\mathrm{B}}}
\newcommand\hf{\hat{H}_{\mathrm{F}}}
\newcommand\hbf{\hat{V}_{\mathrm{BF}}}
\newcommand\hff{\hat{V}_{\nu\pi}}

\newcommand\db{\beta\beta}

\newcommand\gae{g_{\mathrm{A,eff}}}

\newcommand\ga{g_{\mathrm{A}}}
\newcommand\gv{g_{\mathrm{V}}}

\newcommand\ft{\log{ft}}
\newcommand\btm{\beta^{-}}
\newcommand\btp{\beta^{+}}

\newcommand\vd{v_{\mathrm{d}}}

\newcommand\vt{v_{\mathrm{t}}}

\begin{document}

\title{Simultaneous description of $\beta$ decay and low-lying structure of neutron-rich even- and odd-mass Rh and Pd nuclei}

\author{K. Nomura}
\email{knomura@phy.hr}

\author{L. Lotina}
\affiliation{Department of Physics, Faculty of Science, 
University of Zagreb, HR-10000 Zagreb, Croatia}

\author{R.~Rodr\'iguez-Guzm\'an}
\affiliation{
Departamento de F\'isica Aplicada I, Escuela Polit\'ecnica Superior, Universidad de Sevilla, 
Seville, E-41011, Spain}

\author{L.~M.~Robledo}
\affiliation{Departamento de F\'\i sica Te\'orica and CIAFF, Universidad
Aut\'onoma de Madrid, E-28049 Madrid, Spain}

\affiliation{Center for Computational Simulation,
Universidad Polit\'ecnica de Madrid,
Campus de Montegancedo, Bohadilla del Monte, E-28660-Madrid, Spain
}

\date{\today}

\begin{abstract}
The low-energy structure and $\beta$ decay properties of neutron-rich 
even- and odd-mass Pd and Rh nuclei are studied
using a mapping  framework based on the 
nuclear density functional theory 
and the particle-boson coupling scheme.
Constrained 
Hartree-Fock-Bogoliubov calculations using 
the Gogny-D1M energy density functional are 
performed to obtain microscopic inputs 
to determine the interacting-boson  
Hamiltonian employed to describe
the even-even core Pd nuclei. 
The mean-field calculations also 
provide single-particle energies for the odd systems, 
which are used to determine  
essential ingredients of the particle-boson 
interactions for the odd-nucleon systems, and 
of the Gamow-Teller and Fermi transition operators. 
The potential
energy surfaces obtained for even-even Pd isotopes as well as the
spectroscopic properties for the even- and 
odd-mass systems suggest a  transition 
from prolate deformed to $\gamma$-unstable and 
to nearly-spherical shapes. The predicted $\beta$ decay 
$\log{ft}$ values are shown to be sensitive to the details of 
the wave functions for the parent and daughter nuclei, 
and therefore serve as a stringent test of the employed 
theoretical approach. 
\end{abstract}

\maketitle

\section{Introduction}

Precise measurements and theoretical descriptions 
associated with the low-energy nuclear structure
are crucial to the accurate modeling 
and better understanding of fundamental nuclear processes, 
such as,  $\beta$ and double-$\beta$ 
($\db$) decays intimately connected to stellar nucleosynthesis. 
In this context, the low-energy excitations and decay properties 
of neutron-rich nuclei with mass 
$A \approx 100$ and neutron number $N \approx 60$ 
are of particular interest from both the nuclear structure 
and astrophysical points of view. Those nuclei 
exhibit a rich variety of 
phenomena such as shell evolution, onset of collectivity, 
quantum (shape) phase transitions and shape coexistence. 
They are also involved in the rapid neutron-capture ($r$) 
process responsible for the nucleosynthesis of heavy chemical 
elements in explosive environments.

The $\beta$ decay half-lives 
of heavy neutron-rich nuclei have been 
extensively
measured using radioactive-ion 
beams at major experimental facilities around the world.
For example, the neutron-rich $A \approx 110$ nuclei from Kr to Tc 
\cite{nishimura2011}, and from Rb to Sn \cite{lorusso2015} 
have been studied at the RIBF facility at RIKEN. The $A \approx 90$ 
region from  Se to Zr isotopic chains 
has been studied  at the NSCL at MSU \cite{quinn2012}. 
Moreover, 
several $A \approx 100-110$ nuclei are 
of special interest, including $^{96}$Zr, $^{96}$Mo, 
$^{100}$Mo, $^{100}$Ru, $^{110}$Pd, and $^{110}$Cd, 
since they correspond to the parent or daughter 
nuclei for the possible neutrinoless $\db$ 
decays \cite{avignone2008}.

From a theoretical point of view, the consistent 
description of both low-lying nuclear states and $\beta$ 
decay properties represents a major challenge.
Theoretical studies of the 
$\beta$ decay process have been carried out 
within the interacting boson model (IBM)
\cite{navratil1988,DELLAGIACOMA1989,yoshida2002,brant2004,yoshida2013,mardones2016,nomura2020beta-1,nomura2020beta-2,ferretti2020,nomura2022beta}, 
the quasiparticle random-phase approximation (QRPA) 
\cite{alvarez2004,sarriguren2015,boillos2015,pirinen2015,simkovic2013,mustonen2016,suhonen2017,ravlic2021,minato2021}, 
and the large-scale shell model (LSSM) 
\cite{langanke2003,caurier2005,syoshida2018,suzuki2018,akumar2020}. 
The calculation of $\beta$ decay properties 
serves as a stringent test of a given theoretical 
approach, since the decay rate of this process is very sensitive 
to the structure of the wave functions corresponding to the low-energy 
states of both the parent and daughter nuclei. 

In this paper, we present a simultaneous description of 
the low-energy collective excitations and 
$\beta$-decay properties of  even- and 
odd-$A$ neutron-rich Pd and Rh isotopes in the mass range $A \approx 100-120$. 
They represent a region of interest for future experiments and for astrophysical
applications.
Calculations 
are performed within a theoretical 
framework based on the nuclear density functional 
theory and the particle-core coupling scheme. In it even-even nuclei are 
described using the IBM \cite{IBM}. The particle-core 
couplings for 
the odd-mass, and odd-odd nuclei are described using 
the interacting boson-fermion model (IBFM) 
\cite{iachello1979,IBFM} 
and the interacting boson-fermion-fermion 
model (IBFFM) \cite{IBFM,brant1984}, respectively. 
The bosonic-core Hamiltonian is built using microscopic 
input from self-consistent Hartree-Fock-Bogoliubov (HFB) \cite{RS} 
calculations based on the parametrization D1M \cite{D1M}
of the Gogny energy density functional (EDF) \cite{Gogny,robledo2019}.
Essential building blocks of the particle-boson interactions 
and of the Gamow-Teller (GT) and Fermi (F) transition 
operators for the $\beta$ decay are also determined with 
the aid of the same Gogny-EDF results. 
The method has already  been applied to study 
the shape evolution and $\beta$ 
decay properties of the odd-$A$ \cite{nomura2020beta-1} and 
even-$A$ \cite{nomura2020beta-2} nuclei in 
the mass $A \approx 130$ region.  
It has also been employed to study 
even- and odd-$A$ As and Ge nuclei in 
the  $A \approx 70$-80 region 
using microscopic input from  
relativistic Hartree-Bogoliubov calculations, based on 
the density-dependent point-coupling interaction 
\cite{nomura2022beta}.

The main goal of this work is to  examine the performance 
of the method mentioned above in the case of 
neutron-rich nuclei, including 
those for which experimental information is scarce. The results
to be discussed latter on in the paper also illustrate the 
predictive power of the EDF-based IBM  
to describe
the low-lying structure and $\beta$ decay in this region of the nuclear chart
where future experiments are expected.
To identify the relevance of the low-lying 
structures of individual nuclei in the $\beta$ decay, 
we perform a detailed analysis of the wave functions obtained for both 
the parent and daughter nuclei of the decay. 
In addition, we perform 
conventional IBM calculations, with the parameters
for the even-even boson core Hamiltonians 
taken from the earlier phenomenological 
calculation \cite{vanisacker1980}. 
The corresponding results are compared with 
those from the EDF-based IBM calculations. 
Note that the 
present study is restricted to both types 
of allowed $\beta$ decays, i.e., the transition 
conserves parity and
takes place between states that differ in the total angular 
momentum $I$ by $\Delta I=0$ or 1.

To support our choice we note that, like other 
nonrelativistic   
\cite{bender2003} and relativistic 
\cite{vretenar2005,niksic2011} EDFs, 
theoretical approaches  based on the 
parametrizations D1M and D1S \cite{D1S} 
of the Gogny-EDF both at the 
mean-field level and beyond have been extensively 
employed to study the low-energy 
nuclear structure and dynamics
in various regions of the nuclear chart as well as 
fundamental 
nuclear processes 
(see Ref.~\cite{robledo2019} for a 
review, and references therein). 
In particular spectroscopic studies involving collective degrees of
freedom have been carried 
out within the 
symmetry-projected 
generator coordinate method (GCM) \cite{RS} using 
the Gogny forces and involving different levels of sophistication 
\cite{borrajo2016,robledo2019,garrett2020,siciliano2020,siciliano2021,rayner2020oct,rayner2021oct,rayner2022oct}.
Furthermore, the mapping procedure leading 
to an IBM Hamiltonian from microscopic 
Gogny mean-field input has already shown its ability to 
describe spectroscopic properties associated with
shape phase transitions, shape coexistence, 
and octupole deformations in nuclei 
\cite{nomura2012sc,nomura2013hg,nomura2016zr,nomura2015,nomura2020oct,nomura2021oct-u,nomura2021oct-ba,nomura2021oct-zn}.

The paper is organized as follows. The theoretical framework 
is briefly outlined in Sec.~\ref{sec:theory}. 
The excitation spectra 
and electromagnetic transition properties obtained for 
even-even Pd (Sec.~\ref{sec:ee}), 
odd-$A$ Pd and Rh (Sec.~\ref{sec:odd}), 
and odd-odd Rh nuclei (Sec.~\ref{sec:doo}) are discussed. 
The computed $\ft$ values for the $\beta$ decays 
of the odd- and even-$A$ Rh into Pd nuclei 
are discussed in detail in Sec.~\ref{sec:beta}. Finally, Sec.~\ref{sec:summary}
is devoted to the concluding remarks.


\section{Theoretical framework\label{sec:theory}}


In this section, we describe  
the particle-core Hamiltonian 
(Sec.~\ref{sec:par-core-Hamil}), and the 
procedure to build it 
(Sec.~\ref{sec:build-Hamil}). 
Electromagnetic transition operators
are discussed in Sec.~\ref{sec:trans-op}, 
and Gamow-Teller and Fermi operators are 
introduced in Sec.~\ref{sec:GTFtrans-op}.

\subsection{Particle-core Hamiltonian}
\label{sec:par-core-Hamil}

In this study, we use the 
neutron-proton IBM (IBM-2) \cite{OAI,OAIT}. 
In this model both  neutron and proton 
monopole ($s_{\nu}$ and $s_{\pi}$), and 
quadrupole ($d_{\nu}$ and $d_{\pi}$) bosons are considered as fundamental
degrees of freedom. 
From a microscopic point of view \cite{OAIT,OAI}, 
the $s_{\nu}$ ($s_\pi$) and $d_\nu$ ($d_\pi$) bosons 
are associated with the 
collective $S_\nu$ ($S_\pi$) and $D_\nu$ ($D_\pi$) 
pairs of valence neutrons (protons) 
with angular momenta and parity 
$0^{+}$ and $2^{+}$, respectively.
In comparison with the simpler IBM-1, in which 
the neutrons and protons are not distinguished, 
the IBM-2 appears to be more suitable 
to treat $\beta$ decay, since in this 
process both proton and neutron degrees 
of freedom should be explicitly 
taken into account. 
For the model space the neutron $N = $ 50-82 and proton 
$Z = $ 28-50 major  shells  are used. 
Hence for $^{104-124}$Pd, the number of neutron bosons, 
$N_{\nu}$, varies within the range 
$2\leqslant N_\nu\leqslant 8$, while 
the number of the proton bosons is fixed, $N_\pi=2$.

To deal with even-even, odd-mass, 
and odd-odd nuclei on an equal footing, both 
collective and single-particle degrees 
of freedom are treated 
within the framework of 
the neutron-proton IBFFM (IBFFM-2). 
The IBFFM-2 Hamiltonian  reads
\begin{align}
\label{eq:ham-ibffm2}
 \hat{H}=\hb + \hf^{\nu} + \hf^{\pi} + \hbf^{\nu} + \hbf^{\pi} + \hff,
\end{align}
where $\hb$ is the IBM-2 Hamiltonian
representing the 
bosonic even-even core, $\hf^{\nu}$ ($\hf^{\pi}$) 
is the one-body, single-neutron (-proton) 
Hamiltonian, and $\hbf^{\nu}$ ($\hbf^{\pi}$) stands for 
the interaction between the odd neutron (proton) and 
the even-even IBM-2 core. The 
last term $\hff$ represents the residual 
interaction between the odd neutron 
and the odd proton.

The IBM-2 Hamiltonian takes the form
\begin{align}
\label{eq:hb}
 \hb = \epsilon_{d}(\hat{n}_{d_{\nu}}+\hat{n}_{d_{\pi}})
+\kappa\hat{Q}_{\nu}\cdot\hat{Q}_{\pi},
\end{align}
where in the first term, 
$\hat{n}_{d_\rho}=d^\+_\rho\cdot\tilde d_{\rho}$ 
($\rho=\nu$ or $\pi$) is the $d$-boson number operator, 
with $\epsilon_{d}$ the single $d$-boson
energy relative to the $s$-boson one, and 
$\tilde d_{\rho\mu}=(-1)^\mu d_{\rho-\mu}$. 
The second term stands for the quadrupole-quadrupole 
interaction between neutron and proton boson systems 
with strength $\kappa$, and 
$\hat Q_{\rho}=d_{\rho}^\+ s_{\rho} + s_{\rho}^\+\tilde d_{\rho} 
+ \chi_{\rho}(d^\+_{\rho}\times\tilde{d}_{\rho})^{(2)}$ represents the 
bosonic quadrupole operator, with the dimensionless 
parameter $\chi_\rho$.

The single-nucleon Hamiltonian 
$\hf^{\rho}$ takes the form 
\begin{align}
\label{eq:hf}
 \hf^{\rho} = -\sum_{\jr}\epsilon_{\jr}\sqrt{2\jr+1}
  (a_{\jr}^\+\times\tilde a_{\jr})^{(0)},
\end{align}
where $\epsilon_{\jr}$ stands for the 
single-particle energy of the odd neutron $(\rho=\nu)$ 
or proton ($\rho=\pi$) orbital $\jr$. 
$a_{\jr}$ and $a_{\jr}^{\+}$ are
annihilation and creation operators of the 
single particle, respectively. The operator
$\tilde{a}_{\jr}$ is defined as
$\tilde{a}_{\jr\mr}=(-1)^{\jr -\mr}a_{\jr-\mr}$.

In this study, we employ the following 
boson-fermion interaction 
$\hbf^{\rho}$ \cite{IBFM} 
\begin{equation}
\label{eq:hbf}
 \hbf^{\rho}
=\Gamma_{\rho}\hat{V}_{\mathrm{dyn}}^{\rho}
+\Lambda_{\rho}\hat{V}_{\mathrm{exc}}^{\rho}
+A_{\rho}\hat{V}_{\mathrm{mon}}^{\rho}. 
\end{equation}
The first, second, and third 
terms are dynamical quadrupole, 
exchange, and monopole interactions, respectively. 
Within 
the generalized seniority scheme \cite{scholten1985,IBFM}, 
the dynamical and exchange terms 
are assumed to be dominated by the interaction between 
unlike particles. On the other hand, the monopole term 
is assumed to be dominated by the interaction between 
like particles. The explicit form of the different terms in Eq.~(\ref{eq:hbf}) then 
read
\begin{align}
\label{eq:dyn}
&\hat{V}_{\mathrm{dyn}}^{\rho}
=\sum_{\jr\jr'}\gamma_{\jr\jr'}
(a^{\+}_{\jr}\times\tilde{a}_{\jr'})^{(2)}
\cdot\hat{Q}_{\rho'},\\
\label{eq:exc}
&\hat{V}^{\rho}_{\mathrm{exc}}
=-\left(
s_{\rho'}^\+\times\tilde{d}_{\rho'}
\right)^{(2)}
\cdot
\sum_{\jr\jr'\jr''}
\sqrt{\frac{10}{N_{\rho}(2\jr+1)}}
\beta_{\jr\jr'}\beta_{\jr''\jr} \nonumber \\
&{\quad}:\left(
(d_{\rho}^{\+}\times\tilde{a}_{\jr''})^{(\jr)}\times
(a_{\jr'}^{\+}\times\tilde{s}_{\rho})^{(\jr')}
\right)^{(2)}:
+ (\text{H.c.}),\\
\label{eq:mon}
&\hat{V}_{\mathrm{mon}}^{\rho}
=-\hat{n}_{d_{\rho}}
\sum_{\jr}\sqrt{2\jr+1}
  (a_{\jr}^\+\times\tilde a_{\jr})^{(0)},
\end{align} 
where the coefficients 
$\gamma_{\jr\jr'}=(u_{\jr}u_{\jr'}-v_{\jr}v_{\jr'})Q_{\jr\jr'}$, 
and $\beta_{\jr\jr'}=(u_{\jr}v_{\jr'}+v_{\jr}u_{\jr'})Q_{\jr\jr'}$ are
proportional to the matrix elements of the fermion 
quadrupole operator in the single-particle basis
$Q_{\jr\jr'}=\braket{\ell_{\rho}\frac{1}{2}\jr\|Y^{(2)}\|\ell'_\rho\frac{1}{2}\jr'}$.
The operator
$\hat{Q}_{\rho'}$ in Eq.~(\ref{eq:dyn}) is the same boson 
quadrupole operator as in the boson 
Hamiltonian (\ref{eq:hb}). In Eq.~(\ref{eq:exc}) 
the notation $:(\cdots):$ 
stands for normal ordering. 
Within this formalism, the single-particle energy 
$\epsilon_{\jr}$ in Eq.~(\ref{eq:hf}) is replaced with 
the quasiparticle energy $\tilde\epsilon_{\jr}$. 

For the residual 
neutron-proton interaction $\hff$ 
in Eq.~(\ref{eq:ham-ibffm2}), we adopt 
the form \cite{brant1988} 
\begin{align}
\label{eq:hff}
\hff
=4\pi{\vd}
&\delta(\bm{r})
\delta(\bm{r}_{\nu}-r_0)
\delta(\bm{r}_{\pi}-r_0)
\nonumber\\
&+\vt
\left[
\frac{3({\bm\sigma}_{\nu}\cdot{\bf r})
({\bm\sigma}_{\pi}\cdot{\bf r})}{r^2}
-{\bm{\sigma}}_{\nu}
\cdot{\bm{\sigma}}_{\pi}
\right], 
\end{align}
where the first and second terms are surface-delta
and tensor interactions with strength 
parameters $\vd$, and $\vt$, respectively. 
Note that $\bm{r}=\bm{r}_{\nu}-\bm{r}_{\pi}$ 
and $r_0=1.2 A^{1/3}$ fm. 

Table~\ref{tab:comp} summarizes the even-even Pd core 
nuclei, neighboring odd-$A$ Pd and Rh, 
and odd-odd Rh nuclei considered in this study. 
  
\subsection{Procedure to build the Hamiltonian}
\label{sec:build-Hamil}

\begin{figure}[ht]
\begin{center}
\includegraphics[width=\linewidth]{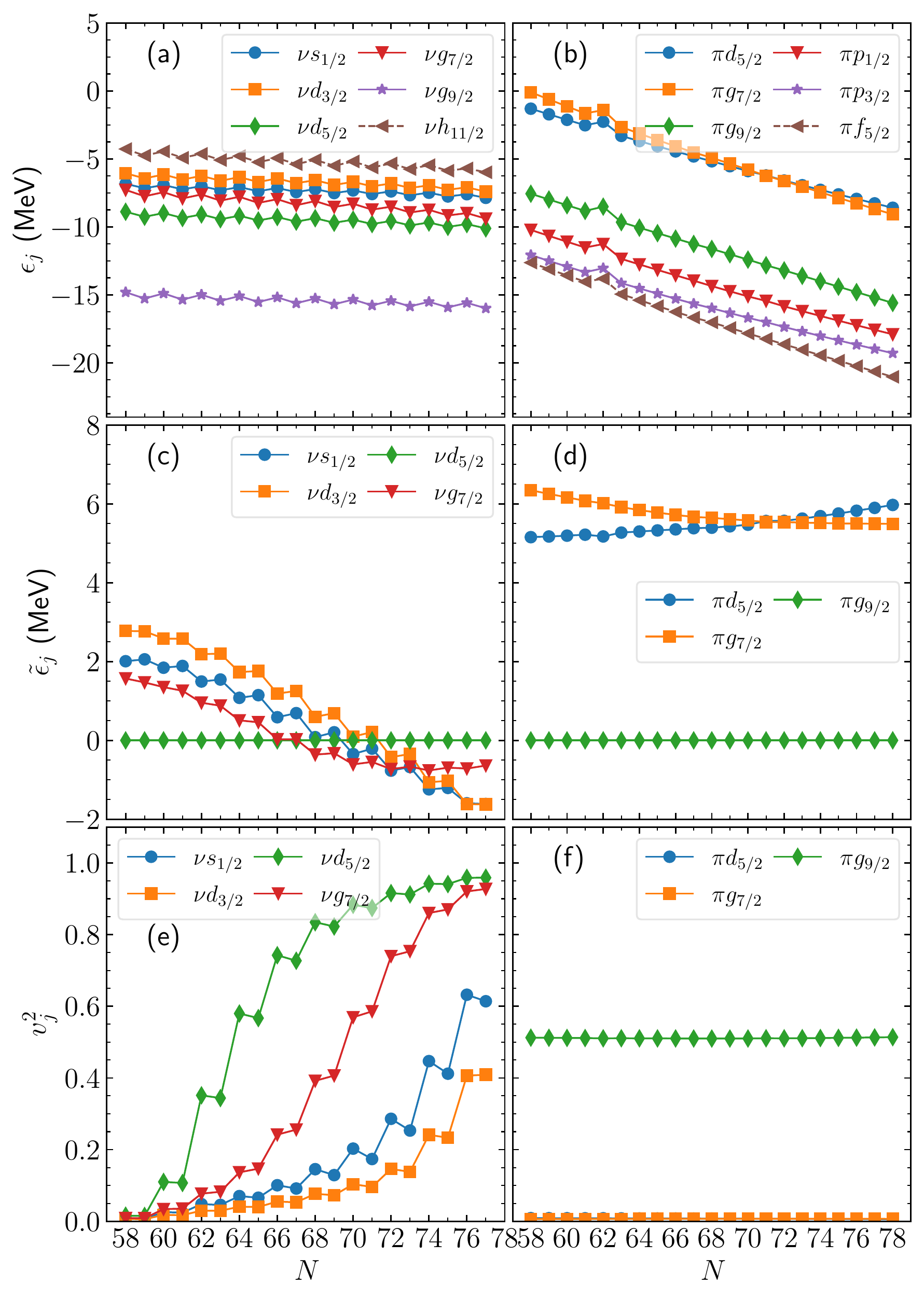}
\caption{(a), (b) The single-particle energies 
$\epsilon_{\jr}$, obtained from the Gogny-D1M HFB 
calculations at the spherical configuration, 
(c), (d) the quasiparticle energies $\tilde\epsilon_{\jr}$, 
and (e), (f) the occupation 
probabilities $v^2_{\jr}$, obtained from 
the BCS calculations. 
Results shown in the left column are 
for the odd neutron 
in the odd-$A$ Pd and even-$A$ Rh nuclei, 
and those in the right column are 
for the odd proton in the even- and odd-$A$ 
Rh nuclei.}
\label{fig:spe}
\end{center}
\end{figure}

\begin{table}
\caption{\label{tab:comp}
Even-even Pd core, and the neighboring odd-$N$ Pd, odd-$Z$ Rh, 
and odd-odd Rh nuclei considered in this study. 
}
 \begin{center}
 \begin{ruledtabular}
  \begin{tabular}{cccc}
even-even core & odd-$N$  & odd-$Z$ & odd-odd  \\
\hline
$^{A}_{46}$Pd$_{N}$ ($58\leqslant N \leqslant 64$) & $^{A+1}_{46}$Pd$_{N+1}$ & $^{A-1}_{45}$Rh$_{N}$ & $^{A}_{45}$Rh$_{N+1}$ \\
[1.0ex]
$^{112}_{46}$Pd$_{66}$ & & $^{111}_{45}$Rh$_{66}$ & \\
[1.0ex]
$^{A}_{46}$Pd$_{N}$ ($68 \leqslant N \leqslant 78$) & $^{A-1}_{46}$Pd$_{N-1}$ & $^{A-1}_{45}$Rh$_{N}$ & $^{A-2}_{45}$Rh$_{N-1}$ \\
  \end{tabular}
 \end{ruledtabular}
 \end{center}
\end{table}

In the initial step a set of 
constrained HFB  calculations for even-even Pd isotopes
based on the parametrization D1M of the 
Gogny-EDF is carried out to obtain the 
microscopic input to build the IBFFM-2 
Hamiltonian. For each 
even-even Pd isotope, those calculations provide 
the corresponding energy surfaces, 
i.e., the total mean-field energies as functions 
of the triaxial quadrupole deformations 
$\beta$ and $\gamma$ \cite{BM}.
For each nucleus, the Gogny-D1M HFB energy surface is 
mapped onto the expectation value of the IBM-2 
Hamiltonian $\hb$ (\ref{eq:hb}) in the boson condensate state 
\cite{ginocchio1980}. This procedure specifies 
the parameters of the boson Hamiltonian, i.e., 
$\epsilon_d$, $\kappa$, $\chi_{\nu}$, and $\chi_{\pi}$. 
For more details about the mapping procedure, 
the reader is referred to 
Refs.~\cite{nomura2008,nomura2010}.

Next, the Hamiltonian $\hf$ of Eq.~(\ref{eq:hf}) 
and the boson-fermion interactions $\hbf$ of  
Eq.~(\ref{eq:hbf}) are determined using the 
procedure of Refs.~\cite{nomura2016odd,nomura2017odd-3}. 
The single-particle energies $\epsilon_{\jr}$ 
of the odd nucleon are obtained from HFB 
calculations constrained to
zero quadrupole deformation. Once the 
single-particle energies are available, 
the quasiparticle energies $\tilde\epsilon_{\jr}$ 
and occupation probabilities $v^2_{\jr}$ are computed 
within the BCS approximation, separately for 
neutron and proton single-particle spaces. 
The empirical
pairing gap $12 A^{-1/2}$ is used. 
We include in the BCS 
calculations the 
$2s_{1/2}$, $1d_{3/2}$, $1d_{5/2}$, 
$0g_{7/2}$, $0g_{9/2}$, and $0h_{11/2}$ 
orbitals for the odd neutron, 
and the $1d_{5/2}$, $0g_{7/2}$, $0g_{9/2}$, 
$2p_{1/2}$, $2p_{3/2}$, and $1f_{5/2}$ 
orbitals for the odd proton. 
The corresponding 
quasiparticle energies 
$\tilde\epsilon_{j_\nu}$ ($\tilde\epsilon_{j_\pi}$), 
and occupation probabilities 
$v^2_{j_\nu}$ ($v^2_{j_\pi}$)
for the odd neutron (proton) 
$2s_{1/2}$, $1d_{3/2}$, $1d_{5/2}$, and 
$0g_{7/2}$ 
($1d_{5/2}$, $0g_{7/2}$, and $0g_{9/2}$)
orbitals are taken as 
the inputs to $\hf^\nu$ ($\hf^\pi$) and $\hbf^\nu$ 
($\hbf^\pi$), respectively.
The strength parameters 
$\Gamma_\rho$, $\Lambda_\rho$, and $A_\rho$ for $\hbf^\rho$ 
are then fixed so that the observed low-energy 
positive-parity levels for the odd-$A$ Pd ($\rho=\nu$) or 
odd-$Z$ Rh ($\rho=\pi$) nuclei are reproduced reasonably 
well.

Finally, the parameters $\vd$ and $\vt$ 
for the residual neutron-proton interaction 
in Eq.~(\ref{eq:hff}) are determined 
\cite{nomura2019dodd} so that the observed low-lying 
positive-parity states for 
each odd-odd Rh nucleus are reasonably well reproduced. 
Note that the same strength parameters as those 
obtained in the previous step for the neighboring 
odd-$A$ nuclei are employed in the IBFFM-2 calculations 
for odd-odd nuclei. On the other hand, the 
quasiparticle energies 
and occupation probabilities of the odd particles 
are independently computed.

Figure~\ref{fig:spe} shows 
the neutron and proton spherical single-particle 
energies ($\epsilon_{j_\nu}$ and $\epsilon_{j_\pi}$), 
resulting from the Gogny-HFB calculations, 
and the quasiparticle energies 
($\tilde\epsilon_{j_\nu}$ and $\tilde\epsilon_{j_\pi}$) 
and occupation probabilities 
($v^2_{j_\nu}$ and $v^2_{j_\pi}$) 
used in the IBFM-2 and IBFFM-2 calculations.

\subsection{Electromagnetic transition operators}
\label{sec:trans-op}

Theories with effective degrees of freedom, like the IBFFM, require the
definition of transition operators to be used in the evaluation of 
electromagnetic transition probabilities. For the electric $E2$ transition
the operator $\hat T^{(E2)}$ to be used  
in the IBFFM-2 takes the form \cite{IBFM}
\begin{align}
 \label{eq:e2}
\hat T^{(E2)}
= \hat T^{(E2)}_\text{B}
+ \hat T^{(E2)}_\text{F} \; ,
\end{align}
where the first and second terms are the 
boson and fermion parts, respectively. 
They are given by
\begin{align}
 \label{eq:e2b}
\hat T^{(E2)}_\mathrm{B}
=\sum_{\rho=\nu,\pi}
e_\rho^\mathrm{B}\hat Q_\rho \; , 
\end{align}
and 
\begin{align}
 \label{eq:e2f}
\hat T^{(E2)}_\mathrm{F}
=-\frac{1}{\sqrt{5}}
&\sum_{\rho=\nu,\pi}
\sum_{\jr\jr'}
(u_{\jr}u_{\jr'}-v_{\jr}v_{\jr'})
\nonumber\\
&\times
\left\langle
\ell_\rho\frac{1}{2}\jr 
\bigg\| 
e^\mathrm{F}_\rho r^2 Y^{(2)} 
\bigg\|
\ell_\rho'\frac{1}{2}\jr'
\right\rangle
(a_{\jr}^\dagger\times\tilde a_{\jr'})^{(2)} \; .
\end{align}

The fixed values  
$e^\mathrm{B}_\nu = e^\mathrm{B}_\pi =0.1$ $e$b 
for the boson effective charges
are taken so that the 
experimental
$B(E2; 2^+_1\rightarrow 0^+_1)$ transition probabilities
are reproduced  for even-even Pd isotopes. The
standard neutron and proton effective 
charges $e^\mathrm{F}_\nu =0.5$ $e$b 
$e^\mathrm{F}_\pi =1.5$ $e$b are employed 
for all the studied odd-nucleon systems. 
The $M1$ transition operator $\hat T^{(M1)}$ is defined as 
\begin{align}
 \label{eq:m1}
\hat T^{(M1)}
=\sqrt{\frac{3}{4\pi}}
&\sum_{\rho=\nu,\pi}
\Biggl[
g_\rho^\mathrm{B}\hat L_\rho
-\frac{1}{\sqrt{3}}
\sum_{\jr\jr'}
(u_{\jr}u_{\jr'}+v_{\jr}v_{\jr'})
\nonumber \\
&\times
\left\langle \jr \| g_l^\rho{\bf l}+g_s^\rho{\bf s} 
\| \jr' \right\rangle
(a_{\jr}^\+\times\tilde a_{\jr'})^{(1)}
\Biggr].
\end{align}
The empirical $g$ factors $g_\nu^\mathrm{B}=0\,\mu_N$ (nuclear 
magneton) and $g_\pi^\mathrm{B}=1.0\,\mu_N$, are adopted
for the neutron and
proton bosons. For the neutron (proton) $g$ factors, the standard 
Schmidt values $g_l^\nu=0\,\mu_N$ and $g_s^\nu=-3.82\,\mu_N$
($g_l^\pi=1.0\,\mu_N$ and $g_s^\pi=5.58\,\mu_N$) 
are used, with $g_s^\rho$ quenched by 30\% 
with respect to the free value.

\subsection{Gamow-Teller and Fermi transition operators}
\label{sec:GTFtrans-op}

As in the electromagnetic case, the transition operators for allowed $\beta${}
decay have to be redefined in terms of the relevant degrees of freedom of
the model. The Gamow-Teller 
$\hat{T}^\mathrm{GT}$ and Fermi $\hat{T}^\mathrm{F}$ 
transition operators take the form
\begin{align}
\label{eq:ogt}
&\hat{T}^{\rm GT}
=\sum_{j_{\nu}j_{\pi}}
\eta_{j_{\nu}j_{\pi}}^{\mathrm{GT}}
\left(\hat P_{j_{\nu}}\times\hat P_{j_{\pi}}\right)^{(1)}, \\
\label{eq:ofe}
&\hat{T}^{\rm F}
=\sum_{j_{\nu}j_{\pi}}
\eta_{j_{\nu}j_{\pi}}^{\mathrm{F}}
\left(\hat P_{j_{\nu}}\times\hat P_{j_{\pi}}\right)^{(0)}, 
\end{align}
with the coefficients 
\begin{align}
\label{eq:etagt}
\eta_{j_{\nu}j_{\pi}}^{\mathrm{GT}}
&= - \frac{1}{\sqrt{3}}
\left\langle
\ell_{\nu}\frac{1}{2}j_{\nu}
\bigg\|{\bm\sigma}\bigg\|
\ell_{\pi}\frac{1}{2}j_{\pi}
\right\rangle
\delta_{\ell_{\nu}\ell_{\pi}},\\
\label{eq:etafe}
\eta_{j_{\nu}j_{\pi}}^{\mathrm{F}}
&=-\sqrt{2j_{\nu}+1}
\delta_{j_{\nu}j_{\pi}}.
\end{align}
In Eqs.~(\ref{eq:ogt}) and (\ref{eq:ofe}), 
$\hat P_{\jr}$ 
represents one of the one-particle creation 
operators 
\begin{subequations}
 \begin{align}
\label{eq:creation1}
&A^{\+}_{\jr\mr} = \zeta_{\jr} a_{{\jr}\mr}^{\+}
 + \sum_{\jr'} \zeta_{\jr\jr'} s^{\+}_\rho (\tilde{d}_{\rho}\times a_{\jr'}^{\+})^{(\jr)}_{\mr}
\\
\label{eq:creation2}
&B^{\+}_{\jr\mr}
=\theta_{\jr} s^{\+}_\rho\tilde{a}_{\jr\mr}
 + \sum_{\jr'} \theta_{\jr\jr'} (d^{\+}_{\rho}\times\tilde{a}_{\jr'})^{(\jr)}_{\mr},
\end{align}
and the annihilation operators
\begin{align}
\label{eq:annihilation1}
&\tilde{A}_{\jr\mr}=(-1)^{\jr-\mr}A_{\jr-\mr}\\ 
\label{eq:annihilation2}
&\tilde{B}_{\jr\mr}=(-1)^{\jr-\mr}B_{\jr-\mr}.  
\end{align}
\end{subequations}
The operators in Eqs.~(\ref{eq:creation1}) 
and (\ref{eq:annihilation1}) 
conserve the boson number, whereas those 
in Eqs.~(\ref{eq:creation2}) and (\ref{eq:annihilation2}) 
do not. 
The operators $\hat{T}^{\rm GT}$ and $\hat{T}^{\rm F}$
are expressed as a combination of two 
of the operators 
in Eqs.~(\ref{eq:creation1})-(\ref{eq:annihilation2}), 
depending on the type of the $\beta$ decay 
studied (i.e., $\beta^+$ or $\btm$) and 
on the particle or hole nature of the valence 
nucleons. In the present case, 
\begin{align}
\hat P_{j_\nu} = 
\left\{
\begin{array}{cc}
\tilde B_{j_\nu,m_\nu} & (N \leqslant 66) \\
\tilde A^\+_{j_\nu,m_\nu} & (N \geqslant 68) \\
\end{array}
\right.
\end{align}
for the $\btm$ decay of the odd-$A$ Rh, while 
\begin{align}
\hat P_{j_\nu} = 
\left\{
\begin{array}{cr}
\tilde A_{j_\nu,m_\nu} & (N \leqslant 65) \\
\tilde B^\+_{j_\nu,m_\nu} & (N \geqslant 67) \\
\end{array}
\right.
\end{align}
for the $\btm$ decay of the even-$A$ Rh. 
On the other hand, $\hat P_{j_\pi}=\tilde A_{j_\pi,m_\pi}$ 
for all the considered $\btm$ decays. Note, that 
Eqs.~(\ref{eq:creation1})--(\ref{eq:annihilation2}) 
are simplified forms of 
the most general one-particle transfer operators 
in the IBFM-2 \cite{IBFM}.

By using the generalized seniority scheme, 
the coefficients $\zeta_{j}$, $\zeta_{jj'}$, 
$\theta_{j}$, and $\theta_{jj'}$ 
in Eqs.~(\ref{eq:creation1}) and (\ref{eq:creation2}) 
can be written as  \cite{dellagiacoma1988phdthesis}
\begin{subequations}
 \begin{align}
\label{eq:zeta1}
\zeta_{\jr}&= 
u_{\jr} \frac{1}{K_{\jr}'}, \\
\label{eq:zeta2}
\zeta_{\jr\jr'}
&= -v_{\jr} 
\beta_{\jr'\jr}
\sqrt{\frac{10}{N_{\rho}(2\jr+1)}}\frac{1}{K K_{\jr}'} , \\ 
\label{eq:theta1}
\theta_{\jr}
&= \frac{v_{\jr}}{\sqrt{N_{\rho}}} 
\frac{1}{K_{\jr}''},\\
\label{eq:theta2}
\theta_{\jr\jr'}
&= u_{\jr} 
\beta_{\jr'\jr}
\sqrt{\frac{10}{2\jr+1}} \frac{1}{K K_{\jr}''}. 
\end{align}
\end{subequations}
The factors $K$, $K_{\jr}'$, and $K_{\jr}''$ 
are defined as
\begin{subequations} 
\begin{align}
\label{eq:k1}
&K = \left( \sum_{\jr\jr'} 
\beta_{\jr\jr'}^{2} \right)^{1/2},\\
\label{eq:k2}
&K_{\jr}' = \left[ 1 + 2 
\left(\frac{v_{\jr}}{u_{\jr}}\right)^{2} \frac{\braket{(\hat 
n_{s_\rho}+1)\hat n_{d_\rho}}_{0^+_1}} {N_\rho(2\jr+1)} \frac{\sum_{\jr'} 
\beta_{\jr'\jr}^{2}}{K^{2}} \right]^{1/2} ,\\
\label{eq:k3}
&K_{\jr}'' = \left[ 
\frac{\braket{\hat n_{s_\rho}}_{0^+_1}}{N_\rho} 
+2\left(\frac{u_{\jr}}{v_{\jr}}\right)^{2} \frac{\braket{\hat 
n_{d_\rho}}_{0^+_1}}{2\jr+1} \frac{\sum_{\jr'} \beta_{\jr'\jr}^{2}}{K^{2}} 
\right]^{1/2},
\end{align} 
\end{subequations}
where $\hat n_{s_{\rho}}$ is the number operator 
for the $s_\rho$ boson and $\braket{\cdots}_{0^+_1}$ 
stands for the expectation 
value of a given operator in the $0^+_1$ ground state 
of the even-even nucleus.
The amplitudes 
$v_{\jr}$ and $u_{\jr}$ 
appearing in Eqs.~(\ref{eq:zeta1})-(\ref{eq:theta2}) and 
(\ref{eq:k1})-(\ref{eq:k3}) 
are the same as those used in the IBFM-2 (or IBFFM-2) 
calculations for the odd-mass (or odd-odd) nuclei. 
No additional parameter is introduced for the GT 
and Fermi operators. For a more detailed account on
$\beta$-decay operators within the IBFM-2 or IBFFM-2 
framework, the reader is also referred to 
Refs.~\cite{dellagiacoma1988phdthesis,DELLAGIACOMA1989,IBFM}.

The $\beta$-decay $ft$ values are given by
\begin{align}
\label{eq:ft}
ft=\frac{K}{|\mfbeta|^2+\left(\frac{\ga}{\gv}\right)^2|\mgtbeta|^2},
\end{align}
where the numeric constant $K$ takes the value $K=6163$~s. The quantities 
$\mfbeta$ and $\mgtbeta$ are the reduced matrix elements 
of the operators $\hat{T}^{\mathrm{F}}$ of Eq.~(\ref{eq:ofe}) 
and $\hat{T}^{\mathrm{GT}}$ of Eq.~(\ref{eq:ogt}), respectively.
Here $\gv$ and $\ga$ are the vector and axial-vector 
coupling constants, respectively. 
In this study, we use the free nucleon values, 
$\gv=1$ and $\ga=1.27$, for the $\beta$ decays of both 
even- and odd-$A$ Rh.

\begin{figure*}[ht]
\begin{center}
\includegraphics[width=.48\linewidth]{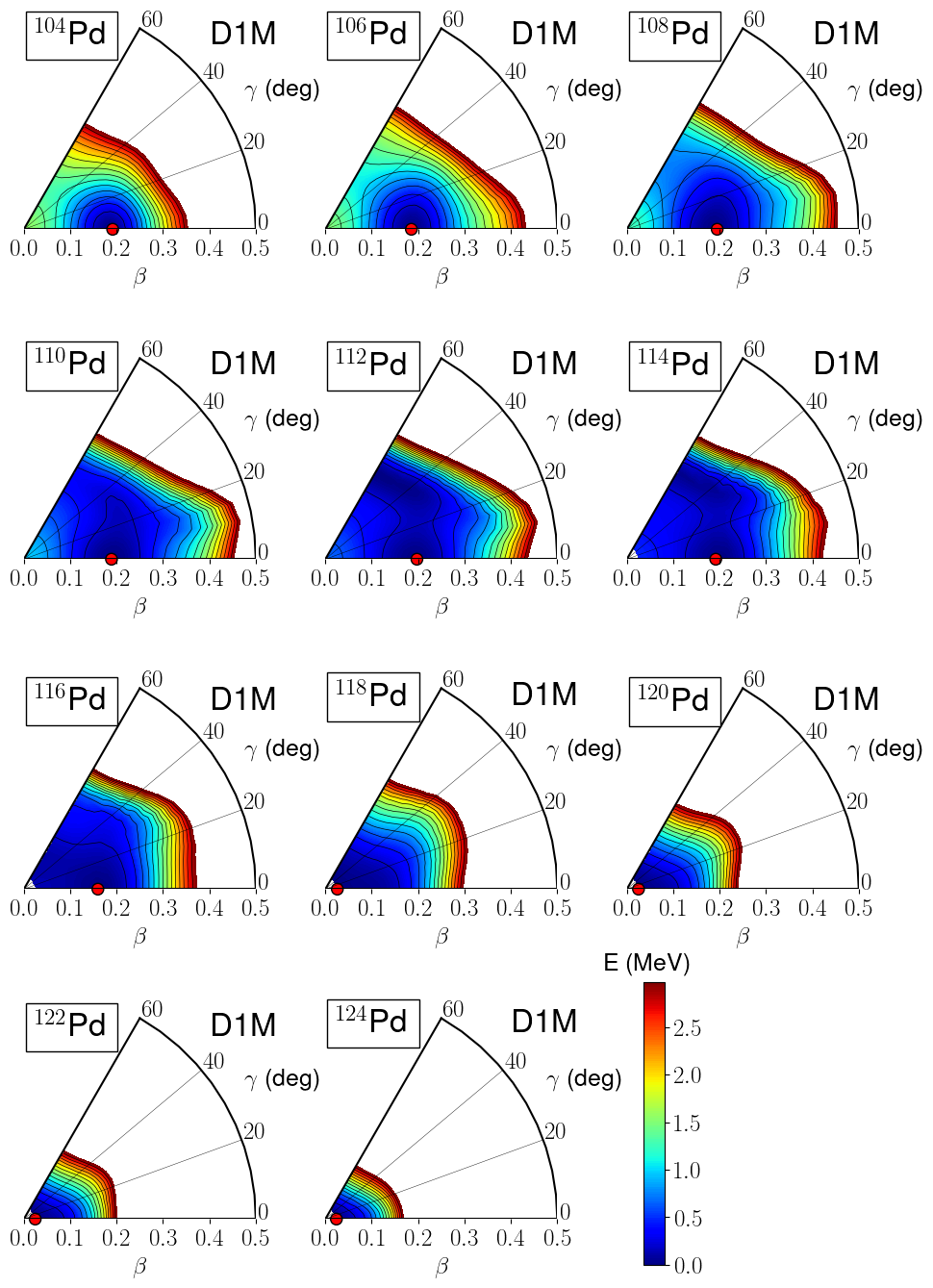}
\includegraphics[width=.48\linewidth]{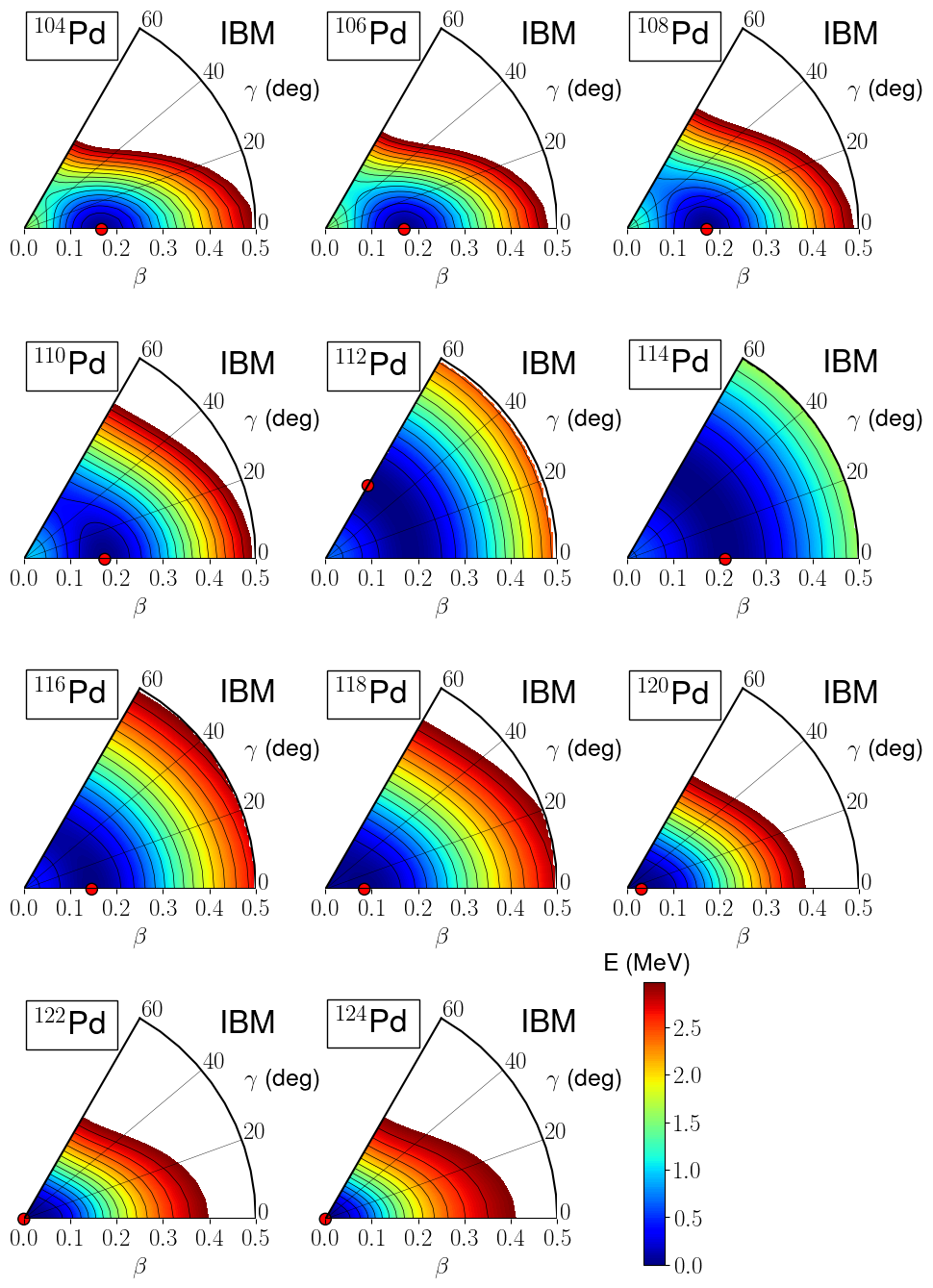}
\caption{The Gogny-D1M HFB and mapped IBM-2 potential energy surfaces
as functions of the $(\beta,\gamma)$ 
deformation parameters for the 
even-even $^{104-124}$Pd nuclei. 
The energy difference between 
neighboring contours is 200 keV. The global minimum 
is identified by a solid circle. 
}
\label{fig:pes}
\end{center}
\end{figure*}


\section{Even-even nuclei\label{sec:ee}}


\subsection{Potential energy surfaces\label{sec:pes}}

The Gogny-D1M HFB and mapped IBM-2 potential energy surfaces
are shown in Fig.~\ref{fig:pes}
as functions of the $(\beta,\gamma)$ 
deformation parameters for the 
even-even $^{104-124}$Pd nuclei. The 
variation of the HFB potential energy surfaces 
as  functions of the neutron number suggests
a transition from prolate (for $N \lesssim 62$) 
to $\gamma$-soft ($64 \lesssim N \lesssim 70$), 
and to nearly spherical  ($N \gtrsim 72$) shapes. 
In particular, both $^{112,114}$Pd exhibit rather flat
potential energy surfaces along the 
$\gamma$ direction. This is what is expected 
in the $\gamma$-unstable O(6) limit of the IBM 
\cite{IBM}. In the case 
of $^{116}$Pd, a flat-bottomed potential with a weak 
$\gamma$ dependence, characteristic 
of the E(5) critical-point symmetry \cite{iachello2000}, 
is obtained.

For each of the considered nuclei, the Gogny-HFB and 
IBM-2 energy surfaces display a similar 
topology in the 
neighborhood of the global minimum (the 
location of the minimum, and the softness in 
the $\beta$ and $\gamma$ directions are similar).
However, the mapped IBM-2 surfaces 
generally become flat at large $\beta$ deformation 
($\beta \gtrsim 0.4$). This difference is 
a consequence of the fact that in the HFB approach
all nucleonic degrees of 
freedom are taken into account while 
the IBM-2 is built on the more limited 
model (valence) space of nucleon pairs. 
However, since the mean-field configurations most relevant 
to the low-energy collective excitations are those 
in the vicinity of the global minimum, the mapping is 
considered specifically in that region 
\cite{nomura2008,nomura2010}.

%
\begin{figure}
\begin{center}
\includegraphics[width=\linewidth]{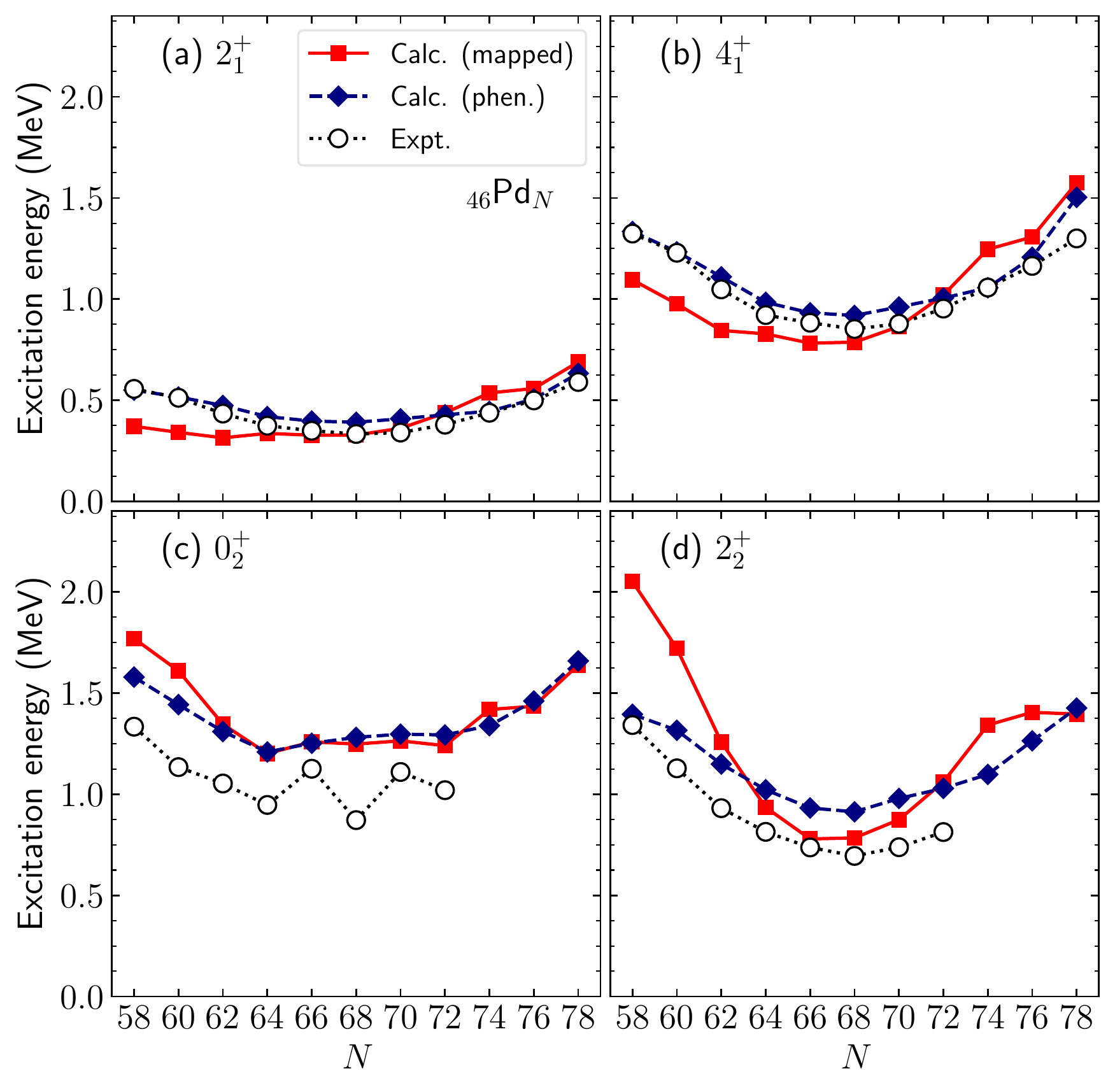}
\caption{Excitation energies 
of the (a) $2^+_1$, (b) $4^+_1$, (c) $0^+_2$, 
and (d) $2^+_2$ (d) states in the even-even $^{104-124}$Pd nuclei.
Results are obtained within the  mapped and phenomenological (phen.) 
IBM-2. Experimental data are taken from Ref.~\cite{data}.}
\label{fig:level-ee}
\end{center}
\end{figure}

\begin{figure}[ht]
\begin{center}
\includegraphics[width=\linewidth]{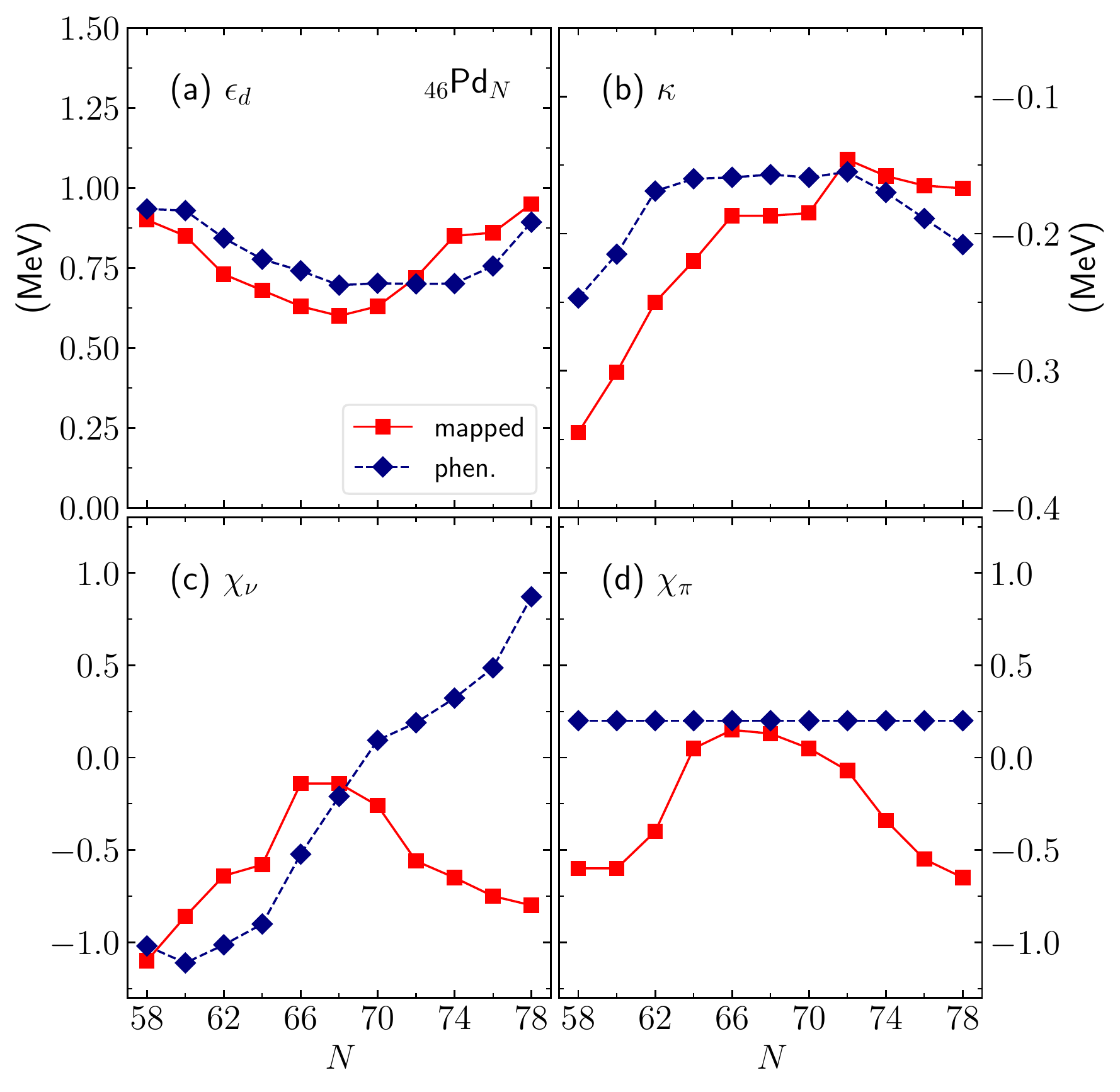}
\caption{Parameters for the even-even boson-core 
Hamiltonian (\ref{eq:hb}) employed in the 
mapped and phenomenological (phen.) 
IBM-2 calculations for  even-even 
Pd isotopes.}
\label{fig:para-ee}
\end{center}
\end{figure}

%
\begin{figure}
\begin{center}
\includegraphics[width=\linewidth]{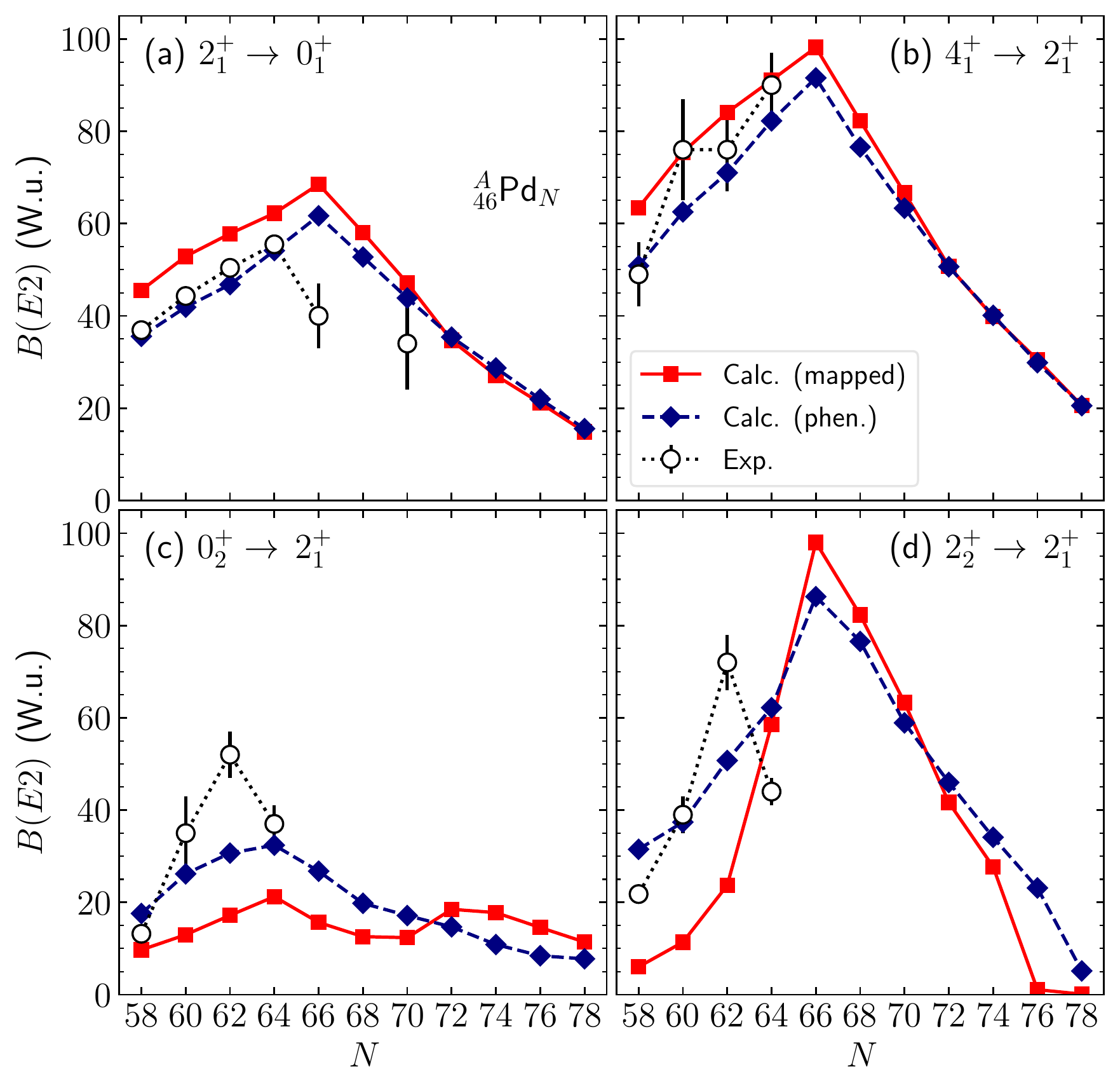}
\caption{The reduced transition probabilities
$B(E2)$ for the transitions (a) $2^+_1\to0^+_1$, 
(b) $4^+_1\to2^+_1$, (c) $0^+_2\to2^+_1$, and (d) $2^+_2\to2^+_1$ 
in even-even Pd isotopes in comparison with the 
experimental data \cite{data}}
\label{fig:e2-ee}
\end{center}
\end{figure}

\subsection{Spectroscopic properties}

The mapped IBM-2 excitation energies 
of the $2^+_1$, $4^+_1$, $0^+_2$, and $2^+_2$ states in 
the even-even $^{104-124}$Pd nuclei are shown 
in Fig.~\ref{fig:level-ee} as functions of 
the neutron number $N$. 
Results obtained using the
conventional IBM-2 approach 
(hereinafter referred to as phenomenological IBM-2), 
with parameters adopted 
from the earlier phenomenological 
study \cite{vanisacker1980}, 
are also included in the plot.
As can be seen from 
the figure, the excitation energies 
decrease toward the middle of the major 
shell, i.e., $N=66$. For 
$N\leqslant 64$, the mapped IBM-2  
$2^+_1$ and $4^+_1$ excitation energies 
underestimate the experimental ones
while the energies of the 
non-yrast  $0^+_2$ and $2^+_2$
states are overestimated. 
In the mapped (phenomenological) IBM-2 
approach the 
ratios $R_{4/2}$  of the 
$4^+_1$ to $2^+_1$ excitation energies are 
2.96 (2.43), 2.86 (2.39), and 2.69 (2.34) 
for $^{104}$Pd, $^{106}$Pd, and $^{108}$Pd, 
respectively. These values should be 
compared with the experimental ratios
of 2.38, 2.40, and 2.41. Thus, the 
mapped  IBM-2 provides excitation 
spectra which are more rotational in character 
than the phenomenological IBM-2 and experimental ones.
Around the neutron midshell $N=66$, both the 
predicted and experimental $2^+_2$ levels 
have the lowest energies, being even below the $4^+_1$ 
state. The $2^+_2$ state is  the bandhead 
of the quasi-$\gamma$ band, and the lowering of 
this state reflects an emergence of pronounced 
$\gamma$ softness.

The IBM-2 parameters obtained for the even-even Pd isotopes 
from the mapping procedure, 
and those determined phenomenologically are shown 
in Fig.~\ref{fig:para-ee}. The phenomenological 
IBM-2 parameters are extracted 
from  earlier fitting calculations for  Pd 
and Ru isotopes \cite{vanisacker1980}. In Ref.~\cite{vanisacker1980}, 
in addition to the terms that appear in Eq.~(\ref{eq:hb}), 
the like-boson interactions, and the so-called 
Majorana terms were included in the model Hamiltonian. 
These terms were, however, shown to 
play a minor role \cite{vanisacker1980}, and are 
omitted in the present study. From Fig.~\ref{fig:para-ee}, 
one sees that
the single-$d$ 
boson energy $\epsilon_d$ and the strength 
$\kappa$ have similar nucleon-number dependence 
for both the mapped and phenomenological
IBM-2 models. A notable quantitative difference is that the 
derived $\kappa$ values for the former are 
$\approx$ 1.4 larger in magnitude  than for the latter. 
The behavior of the parameter $\chi_\nu$ is 
different in the two approaches
for $N \geqslant 70$.
The sign and absolute value of 
the sum $\chi_\nu + \chi_\pi$ reflect the extent 
of $\gamma$ softness and whether the nucleus is prolate 
or oblate deformed. In both  calculations, 
the sum is negative, $\chi_\nu + \chi_\pi <0$, 
for $N \lesssim 64$, 
indicating prolate deformation, 
and takes nearly vanishing values, 
$\chi_\nu + \chi_\pi \approx 0$, 
around the neutron midshell $N=66$, 
reflecting  $\gamma$ softness.
 However, for $N \geqslant 70$, the sum is negative (positive)
in the mapped (phenomenological) calculations, 
implying  prolate (oblate) deformation. 
Note that a fixed value $\chi_\pi=0.2$ is employed in the 
phenomenological 
IBM-2 calculations, whereas in the mapped approach this 
parameter exhibits a strong nucleon 
number dependence.

The $B(E2)$ transition probabilities, computed
within the mapped and phenomenological 
IBM-2 models, are plotted in  Fig.~\ref{fig:e2-ee}
as functions of the neutron number $N$.
The same $E2$ effective boson charge 
is used for the quadrupole 
operators in the two sets of the IBM-2 calculations. 
The $B(E2;2^+_1 \to 0^+_1)$ and $B(E2;4^+_1 \to 2^+_1)$ 
values obtained in the mapped IBM-2 calculations
agree reasonably well with the experiment, exception made 
of $^{112}$Pd. Both the mapped
and phenomenological IBM-2 calculations predict 
$B(E2;0^+_2 \to 2^+_1)$ and $B(E2;2^+_2 \to 2^+_1)$ 
rates with  similar trends as functions of 
$N$.  However, the  mapped IBM-2
scheme provides smaller $B(E2;0^+_2 \to 2^+_1)$ values
for Pd isotopes with $58 \leqslant N \leqslant 62$.
The enhancement of the predicted
$B(E2;2^+_2 \to 2^+_1)$ transition rates around the midshell 
$N=66$ [see Fig.~\ref{fig:e2-ee}(d)] can be considered 
as another signature of  $\gamma$ soft deformation.

%
\begin{figure*}
\begin{center}
\includegraphics[width=.8\linewidth]{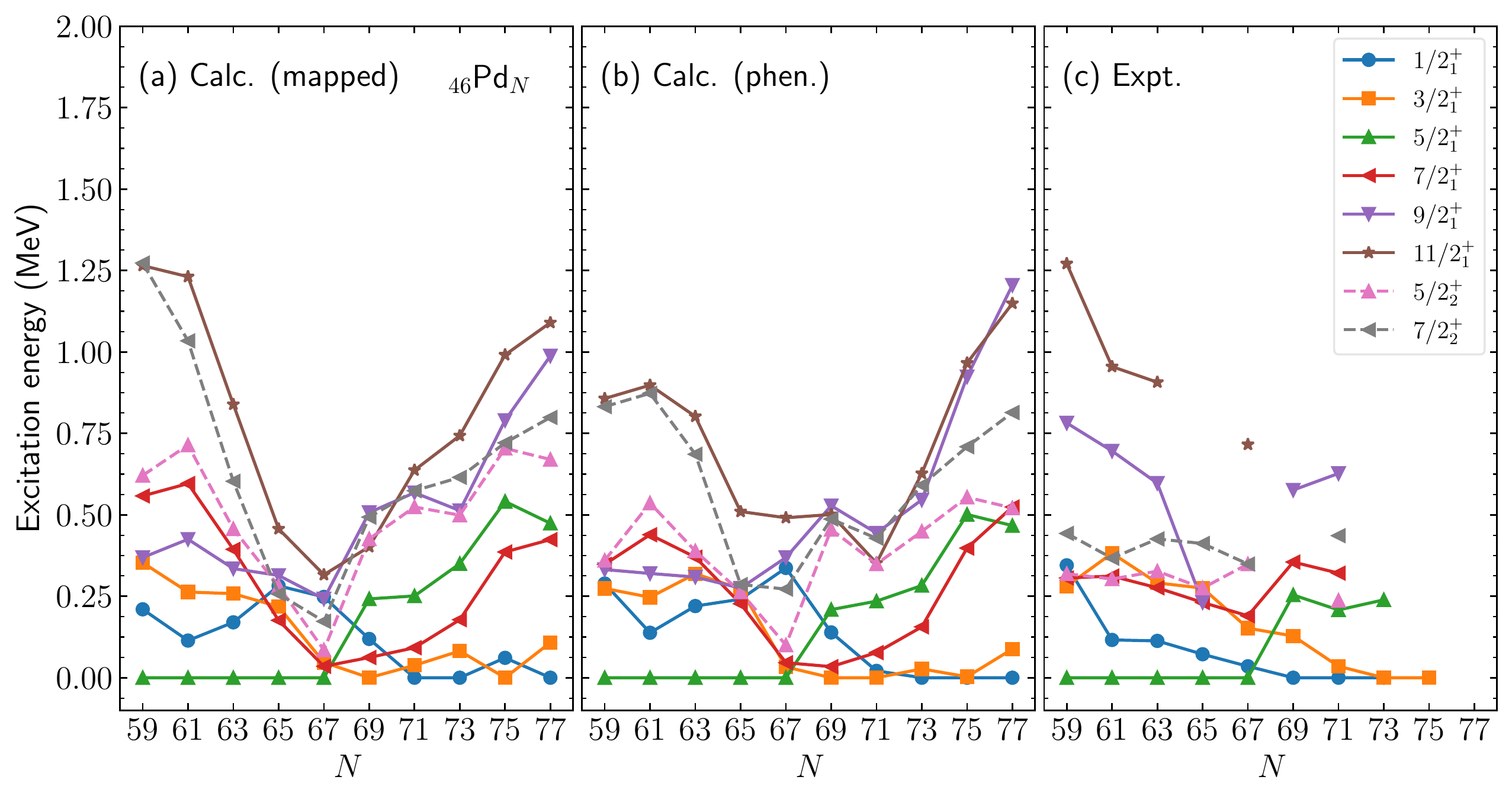}
\caption{Excitation energies of the low-lying 
positive-parity states 
obtained for odd-$A$ Pd isotopes within the 
IBFM-2 with the boson-core Hamiltonian 
determined by (a) mapping the Gogny-D1M EDF, and (b) using
the phenomenological fit. The experimental data 
included in panel (c) are taken from 
Ref.~\cite{kurpeta2018-117Pd} 
for $^{117}$Pd, from Ref.~\cite{kurpeta2022-119Pd} 
for $^{119}$Pd, and from the NNDC database \cite{data} 
for the other nuclei.}
\label{fig:oddpd}
\end{center}
\end{figure*}

%
\begin{figure*}
\begin{center}
\includegraphics[width=.8\linewidth]{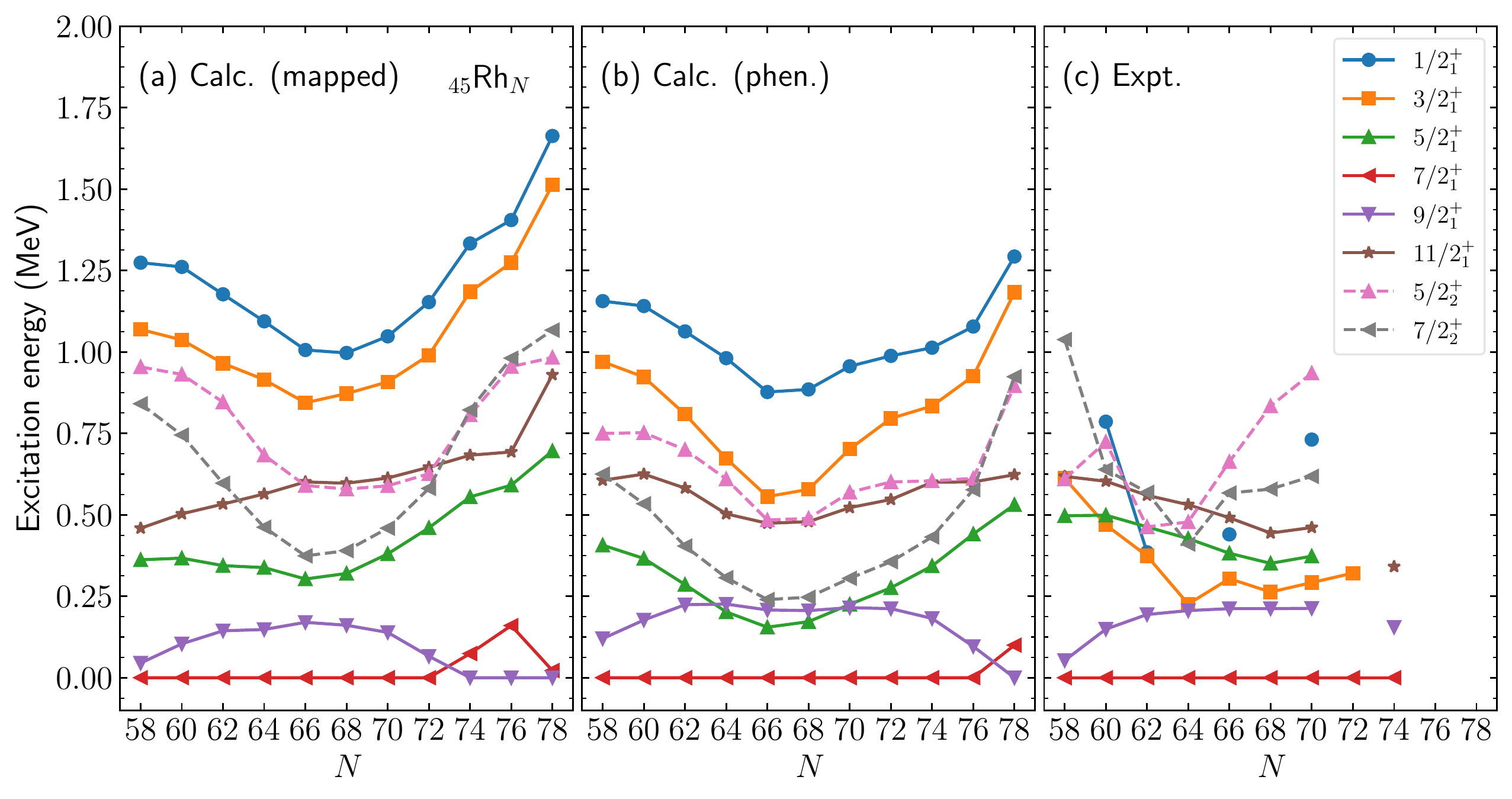}
\caption{The same as Fig.~\ref{fig:oddpd}, but for   
odd-$Z$ Rh nuclei. The experimental data are taken 
from Ref.~\cite{data}.}
\label{fig:oddrh}
\end{center}
\end{figure*}

\begin{figure}[ht]
\begin{center}
\includegraphics[width=\linewidth]{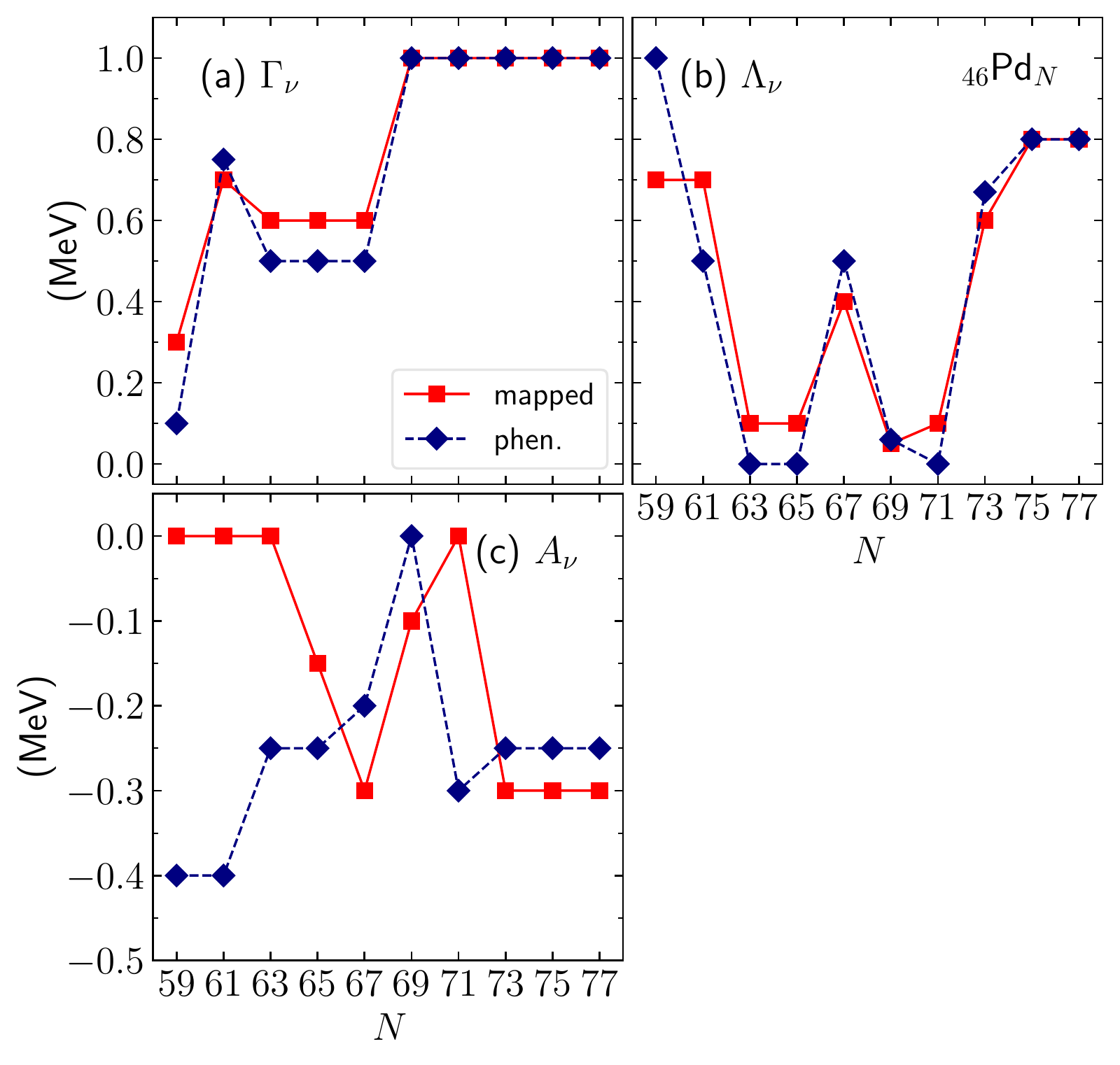}
\caption{Parameters of the mapped and phenomenological IBFM-2 
Hamiltonian (\ref{eq:hb}) for odd-$N$ Pd nuclei.}
\label{fig:para-odd}
\end{center}
\end{figure}

%
\begin{figure}
\begin{center}
\includegraphics[width=\linewidth]{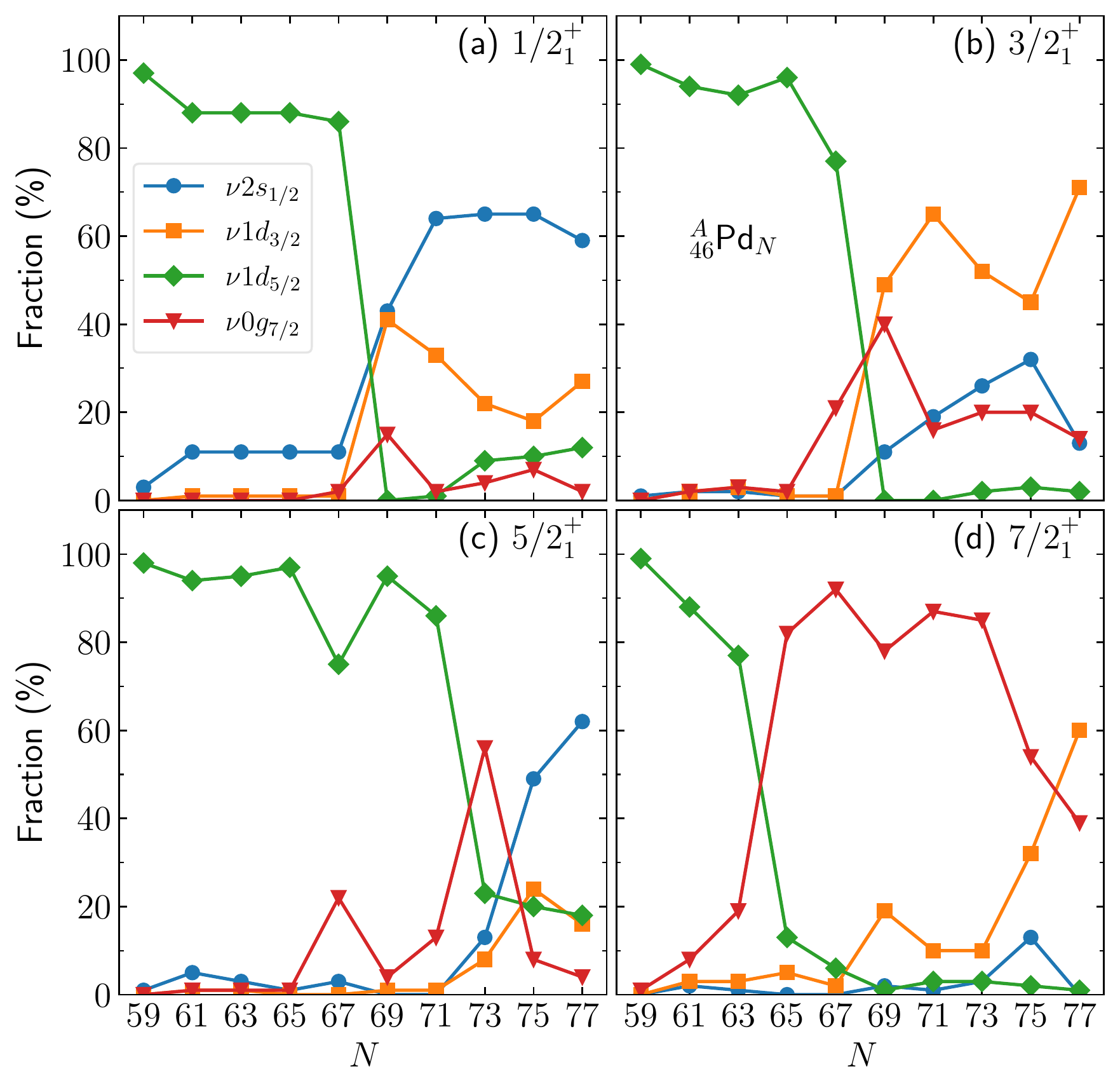}
\caption{Fractions (in percent \%) of the 
neutron $\nu 2s_{1/2}$, $\nu 1d_{3/2}$, $\nu 1d_{5/2}$, 
and $\nu 0g_{7/2}$ single-particle configurations 
in the wave functions for the (a) ${1/2}^+_1$, 
(b) ${3/2}^+_1$, (c) ${5/2}^+_1$, and (d) ${7/2}^+_1$ 
states in odd-$A$ Pd nuclei. 
The wave functions are obtained within the mapped
IBFM-2  scheme based on Gogny-D1M EDF calculations.
}
\label{fig:oddpd-wf}
\end{center}
\end{figure}


\section{Odd-$A$ Pd and Rh nuclei\label{sec:odd}}


The excitation energies of the low-lying 
positive-parity states 
obtained for the odd-$A$ Pd isotopes $^{105-123}$Pd are depicted 
in Fig.~\ref{fig:oddpd}. The results obtained 
within the 
IBFM-2  model with boson-core Hamiltonian 
determined by mapping the Gogny-D1M EDF [Fig.~\ref{fig:oddpd}(a)] 
and the those obtained from 
phenomenological calculations of 
Ref.~\cite{vanisacker1980} [Fig.~\ref{fig:oddpd}(b)] 
are compared with experimental data 
\cite{kurpeta2018-117Pd,kurpeta2022-119Pd,data}.
The two IBFM-2 calculations, using different boson-core 
Hamiltonian parameters, provide an overall consistent 
description of the experimental excitation energies. 
As can be seen from the figure, the experimental 
data display a change in the ground state spin 
from $N=67$ to 69. The corresponding even-even core nuclei, 
$^{114}$Pd and $^{116}$Pd, are in the 
transitional region, for which the 
potential energy surfaces are suggested to be 
considerably $\gamma$ soft (see Fig.~\ref{fig:pes}).
The sudden change in the ground-state spin of the 
odd-$A$ neighbor, therefore, reflects the transition 
that takes place in the even-even core systems 
from the $\gamma$ unstable shape, which is associated 
with an O(6)-like potential, 
to the E(5)-like structure characterized 
by a flat-bottomed potential.

The excitation energies of the low-lying 
positive-parity states 
obtained for the odd-$A$ isotopes $^{103-123}$Rh are depicted 
in Fig.~\ref{fig:oddrh}. Experimentally, the ground 
states of  these isotopes have spin $I^{\pi}={7/2}^+$. 
Exceptions are made of some of the heaviest isotopes, 
and similar 
results are predicted within both the mapped and 
phenomenological calculations. Both theoretically 
and experimentally, 
some of the energy levels exhibit an
approximate parabolic behavior 
with a minimum around 
the middle of 
the major shell, $N \approx 66$. 
For $^{103-123}$Rh, the order of most of the 
energy levels remains 
unchanged in the whole isotopic chain
within both the mapped and 
phenomenological IBFM-2 calculations.
This situation is in a sharp contrast with the one 
in the odd-$A$ Pd (see Fig.~\ref{fig:oddpd}), 
in which the structural 
change along the isotopic chain occurs more rapidly. 
Note that the low-lying states 
of the odd-$A$ Rh nuclei are accounted for almost purely 
by the proton $\pi 0g_{9/2}$ single-particle 
configuration while more than one 
single-particle orbital is considered for the 
odd-$A$ Pd. 
The occupation 
number of the odd proton in the $\pi 0g_{9/2}$ 
orbital is also nearly constant along the whole Rh isotopic 
chain [see Fig.~\ref{fig:spe}(f)], whereas the occupation 
probabilities for the odd neutron in the odd-$A$ Pd 
vary significantly with $N$ [see Fig.~\ref{fig:spe}(e)]. 
Furthermore, as shown below, the strength 
parameters for $\hbf$ are fixed in the case of  odd-$A$ Rh
nuclei while they depend on the  
boson number 
for  odd-$A$ Pd isotopes.

The strength parameters of the 
boson-fermion interaction (\ref{eq:hbf}) 
for odd-$N$ Pd nuclei are 
shown in Fig.~\ref{fig:para-odd}. 
These parameters are chosen 
so that the 
ground-state spin, and energies of a few low-lying 
levels are reproduced reasonably well. The 
parameters for the two IBFM-2 calculations
are rather similar, with an exception made of the 
monopole strength $A_\nu$ 
for $59 \leqslant N \leqslant 63$.
Note that common quasiparticle energies 
$\tilde\epsilon_{\jr}$ and 
occupation probabilities $v^2_{\jr}$ are used for 
both  IBFM-2 calculations.
The parameters 
for the $^{123}$Pd$_{77}$ nucleus, where no experimental 
data are  available, are taken to be the same as those for 
the adjacent nucleus $^{121}$Pd$_{75}$. As can be seen 
from the figure, the IBFM-2 parameters 
turn out to have a strong $N$-dependence that 
reflects the rapid structural change in the 
odd-$A$ Pd isotopes. On the other hand, constant 
strength parameters $\Gamma_\pi=0.6$ (0.0) MeV, 
$\Lambda_{\pi}=0.6$ (0.75) 
MeV, and $A_\pi=0.0$ ($-0.25$) MeV
reproduce reasonably well the experimental data for 
odd-$A$ Rh nuclei in the mapped (phenomenological)
calculations.

%
\begin{table}
\caption{\label{tab:oddpd-em}
$B(E2)$ rates (in Weisskopf units, W.u.), 
quadrupole moment $Q(I)$ (in $e$b), $B(M1)$ rates 
(in W.u. $\times 10^{-3}$), 
and magnetic dipole moments $\mu(I)$ (in $\mu_N$) 
obtained
for odd-$A$ Pd nuclei
within the mapped and phenomenological 
IBFM-2 calculations.  Experimental 
data are taken from Ref.~\cite{data,stone2005}. 
}
 \begin{center}
 \begin{ruledtabular}
  \begin{tabular}{lcccc}
 & & \multicolumn{2}{c}{Calc.} & \\
\cline{3-4}
 & & mapped & phen. & Expt. \\
\hline
$^{105}$Pd
& $B(E2;1/2^+_{1}\to3/2^+_{1})$ & $25$ & $13$ & $2.0^{+91}_{-16}$ \\
& $B(E2;1/2^+_{1}\to5/2^+_{1})$ & $90$ & $45$ & $2.64(15)$ \\
& $B(E2;1/2^+_{2}\to3/2^+_{1})$ & $0.6$ & $3.1$ & $0.9^{+12}_{-7}$ \\
& $B(E2;1/2^+_{2}\to5/2^+_{1})$ & $0.04$ & $0.9$ & $8.4(9)$ \\
& $B(E2;3/2^+_{1}\to5/2^+_{1})$ & $44$ & $40$ & $4.6(7)$ \\
& $B(E2;3/2^+_{1}\to3/2^+_{3})$ & $0.05$ & $5.1$ & $>0.21$ \\
& $B(E2;3/2^+_{3}\to5/2^+_{2})$ & $0.01$ & $2.7$ & $>2.2$ \\
& $B(E2;5/2^+_{1}\to5/2^+_{2})$ & $15$ & $29$ & $1.8(4)$ \\
& $B(E2;7/2^+_{1}\to5/2^+_{1})$ & $24$ & $33$ & $0.30(4)$ \\
& $B(E2;9/2^+_{1}\to5/2^+_{1})$ & $57$ & $40$ & $14.3(13)$ \\
& $B(M1;1/2^+_{1}\to3/2^+_{1})$ & $372$ & $280$ & $14.9^{+20}_{-21}$ \\
& $B(M1;1/2^+_{1}\to1/2^+_{2})$ & $0.93$ & $1.5$ & $7.8(8)$ \\
& $B(M1;1/2^+_{2}\to3/2^+_{1})$ & $37$ & $7$ & $45^{+6}_{-5}$ \\
& $B(M1;3/2^+_{1}\to5/2^+_{1})$ & $31$ & $4.4$ & $20.3(22)$ \\
& $B(M1;3/2^+_{1}\to3/2^+_{3})$ & $0.012$ & $0.0004$ & $>5.9$ \\
& $B(M1;3/2^+_{3}\to5/2^+_{2})$ & $0.0026$ & $2.9$ & $>47$ \\
& $B(M1;5/2^+_{1}\to5/2^+_{2})$ & $13$ & $1.6$ & $19(3)$ \\
& $B(M1;5/2^+_{3}\to3/2^+_{1})$ & $4.7$ & $0.47$ & $>0.40$ \\
& $B(M1;5/2^+_{3}\to7/2^+_{2})$ & $52$ & $32$ & $>25$ \\
& $B(M1;7/2^+_{1}\to5/2^+_{1})$ & $31$ & $3.7$ & $10.6(12)$ \\
& $Q(5/2^+_{1})$ & $-0.54$ & $-0.27$ & $+0.660(11)$ \\
& $\mu(3/2^+_{1})$ & $-0.56$ & $-0.64$ & $-0.074(13)$ \\
& $\mu(5/2^+_{1})$ & $-1.19$ & $-1.32$ & $-0.642(3)$ \\
& $\mu(5/2^+_{2})$ & $-0.67$ & $-0.76$ & $+0.95(20)$ \\
[1.0ex]
$^{107}$Pd
& $B(E2;1/2^+_{1}\to5/2^+_{1})$ & $112$ & $90$ & $0.58(7)$ \\
& $\mu(5/2^+_{1})$ & $-1.06$ & $-1.05$ & $0.735(7)$ \\
[1.0ex]
$^{109}$Pd
& $B(E2;1/2^+_{1}\to5/2^+_{1})$ & $97$ & $76$ & $1.36(18)$ \\
& $B(E2;3/2^+_{1}\to5/2^+_{1})$ & $58$ & $48$ & $8(8)$ \\
& $B(M1;3/2^+_{1}\to5/2^+_{1})$ & $4.4$ & $4.4$ & $2.2(8)$ \\
& $B(M1;5/2^+_{2}\to3/2^+_{1})$ & $159$ & $142$ & $11.7(19)$ \\
& $B(M1;7/2^+_{2}\to5/2^+_{1})$ & $3.2$ & $0.13$ & $3.6(4)$ \\
 \end{tabular}
 \end{ruledtabular}
 \end{center}
\end{table}

%
\begin{table}
\caption{\label{tab:oddrh-em}
The same as in Table~\ref{tab:oddpd-em}, but for 
odd-$A$ Rh nuclei.
}
 \begin{center}
 \begin{ruledtabular}
  \begin{tabular}{lcccc}
 & & \multicolumn{2}{c}{Calc.} & \\
\cline{3-4}
& & mapped & phen. & Expt. \\
\hline
$^{103}$Rh
& $B(E2;5/2^+_{1}\to7/2^+_{1})$ & $33$ & $31$ & $2.0(6)$ \\
& $B(E2;5/2^+_{1}\to9/2^+_{1})$ & $28$ & $13$ & $0.107(33)$ \\
& $B(M1;5/2^+_{1}\to7/2^+_{1})$ & $471$ & $354$ & $40(12)$ \\
& $B(M1;9/2^+_{1}\to7/2^+_{1})$ & $1.0$ & $1.9$ & $43(12)$ \\
& $\mu(7/2^+_{1})$ & $4.85$ & $4.88$ & $+4.540(11)$ \\
& $\mu(9/2^+_{1})$ & $5.69$ & $5.62$ & $+4.9(8)$ \\
[1.0ex]
$^{107}$Rh
& $B(E2;3/2^+_{1}\to7/2^+_{1})$ & $4.73$ & $1.62$ & $0.16(2)$ \\
[1.0ex]
$^{109}$Rh
& $B(E2;3/2^+_{1}\to3/2^+_{2})$ & $0.14$ & $0.18$ & $1.7\times{10}^2(5)$ \\
& $B(E2;3/2^+_{1}\to7/2^+_{1})$ & $4.41$ & $0.01$ & $0.0174(5)$ \\
& $B(E2;3/2^+_{2}\to7/2^+_{1})$ & $5.3$ & $5.9$ & $26.1(19)$ \\
& $B(E2;5/2^+_{1}\to9/2^+_{1})$ & $7.9$ & $5.3$ & $>23$ \\
& $B(E2;5/2^+_{2}\to3/2^+_{1})$ & $12$ & $5.8$ & $1.7(7)$ \\
& $B(E2;5/2^+_{2}\to3/2^+_{2})$ & $22$ & $9$ & $7.E+1(3)$ \\
& $B(E2;7/2^+_{2}\to3/2^+_{1})$ & $8$ & $15$ & $131(12)$ \\
& $B(M1;5/2^+_{1}\to3/2^+_{3})$ & $5.2$ & $8.6$ & $>220$ \\
& $B(M1;5/2^+_{1}\to3/2^+_{1})$ & $818$ & $414$ & $>0.40$ \\
& $B(M1;5/2^+_{2}\to3/2^+_{1})$ & $37$ & $289$ & $2.4(3)$ \\
& $B(M1;5/2^+_{2}\to3/2^+_{2})$ & $152$ & $207$ & $2.2(15)$ \\
& $B(M1;5/2^+_{2}\to3/2^+_{3})$ & $18$ & $210$ & $2.5(4)$ \\
& $B(M1;5/2^+_{2}\to7/2^+_{1})$ & $231$ & $112$ & $4.1{\times}10^{-2}(6)$ \\
& $B(M1;7/2^+_{2}\to9/2^+_{1})$ & $318$ & $611$ & $0.25(6)$ \\
& $B(M1;7/2^+_{1}\to7/2^+_{2})$ & $7.6$ & $8.9$ & $6.6{\times}10^{-2}(8)$ \\
& $B(M1;3/2^+_{1}\to3/2^+_{2})$ & $27$ & $48$ & $0.58(12)$ \\
& $B(M1;3/2^+_{1}\to3/2^+_{3})$ & $158$ & $256$ & $1.18(11)$ \\
& $B(M1;3/2^+_{2}\to3/2^+_{3})$ & $276$ & $32$ & $0.32(10)$ \\
& $B(M1;5/2^+_{1}\to7/2^+_{1})$ & $233$ & $172$ & $>3.2$ \\
& $B(M1;9/2^+_{1}\to7/2^+_{1})$ & $4.3$ & $18$ & $>58$ \\
 \end{tabular}
 \end{ruledtabular}
 \end{center}
\end{table}

Experimental data for  electromagnetic transitions 
and moments are available 
for  odd-$A$ Pd and Rh nuclei with $N \leqslant 65$. 
The
predicted $B(E2)$ and $B(M1)$ transition strengths 
as well as the electric quadrupole $Q(I^\pi)$ and magnetic 
dipole $\mu(I^\pi)$ moments for 
the low-lying positive-parity states in odd-$A$ Pd
are given in  Table~\ref{tab:oddpd-em}. In most of 
the cases, the mapped and phenomenological calculations
provide similar results. Large values are obtained for
the $B(E2;{1/2}^+_1 \to {5/2}^+_1)$ 
(in $^{105}$Pd, $^{107}$Pd and $^{109}$Pd), 
$B(E2;{3/2}^+_1 \to {5/2}^+_1)$ (in $^{105}$Pd and $^{109}$Pd), 
and $B(E2;{9/2}^+_1 \to {5/2}^+_1)$ (in $^{105}$Pd)
transitions. The experimental data, however, suggest 
that these $E2$ transitions are  weaker.
The $B(E2)$ and $B(M1)$ rates 
corresponding to some transitions in 
odd-$A$ Rh nuclei are given in Table~\ref{tab:oddrh-em}. 
The large $B(E2;{5/2}^+_1 \to {7/2}^+_1)$ and 
$B(E2;{5/2}^+_1 \to {9/2}^+_1)$ rates obtained for $^{103}$Rh
overestimate the experimental rates by several orders 
of magnitude.

The deviation of the predicted $B(E2)$ and $B(M1)$ 
transition rates for  odd-$A$ systems 
with respect to the experiment
could be interpreted in terms of 
the structure of the
corresponding IBFM-2 wave functions. The 
components of the IBFM-2 wave functions for 
the  low-lying 
states of odd-$A$ Pd isotopes are shown  in Fig.~\ref{fig:oddpd-wf}.
They are  associated 
with the single(quasi)-particle 
orbitals $\nu 2s_{1/2}$, $\nu 1d_{3/2}$, 
$\nu 1d_{5/2}$, and $\nu 0g_{7/2}$. Only 
components  obtained within 
the mapped framework are shown as illustrative
examples, while qualitatively similar 
results are obtained using the 
phenomenological approach. 
The states
considered for odd-$A$ Rh nuclei are almost purely made 
of the proton $0g_{9/2}$ configuration (with a weight of
$\approx 99$ \%). Therefore, the corresponding 
wave function contents are not shown in the plot.
As can be seen from the figure, the neutron 
$1d_{5/2}$ configuration accounts for most of the IBFM-2 
wave functions for the ${1/2}^+_1$, 
${3/2}^+_1$, ${5/2}^+_1$, and ${7/2}^+_1$ in 
odd-$A$ Pd nuclei with $N \lesssim 67$. 
However, the description of these wave 
functions in both the mapped and phenomenological 
IBFM-2 calculations in the present study 
may not be adequate, and this leads to 
some of the considerable 
disagreements between the calculated 
and experimental 
electromagnetic properties, including the 
$B(E2;{1/2}^+_1 \to {5/2}^+_1)$ values in 
$^{105}$Pd, $^{107}$Pd, and $^{109}$Pd 
(see Table~\ref{tab:oddpd-em}). 
The deficiency of the IBFM-2 wave functions 
could arise from various deficiencies 
of the present model calculations, such as the 
choice of the single-particle space, 
the quasiparticle energies and occupation 
probabilities of the odd particle, and 
the effective charges involved in 
the transition operators, which are kept 
constant for all nuclei. 
On the other hand, 
earlier IBFM-2 fitting calculations 
in the same mass region 
\cite{arias1987,yoshida2002} 
obtained $E2$ and $M1$ properties 
consistent with experiment.

%
\begin{figure*}
\begin{center}
\includegraphics[width=.8\linewidth]{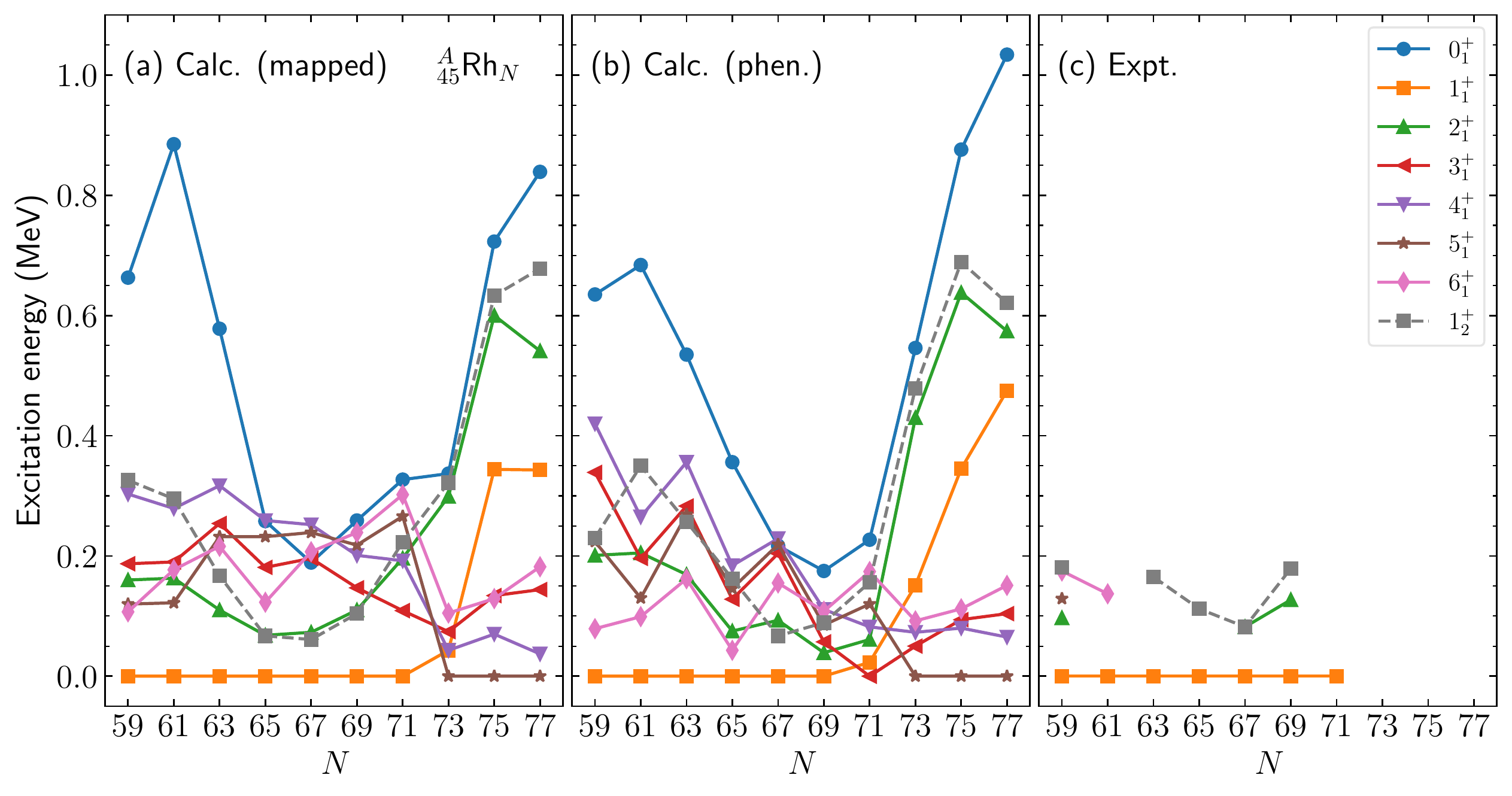}
\caption{The same as Fig.~\ref{fig:oddpd}, but for the odd-odd Rh isotopes.}
\label{fig:doorh}
\end{center}
\end{figure*}

\begin{figure}[ht]
\begin{center}
\includegraphics[width=\linewidth]{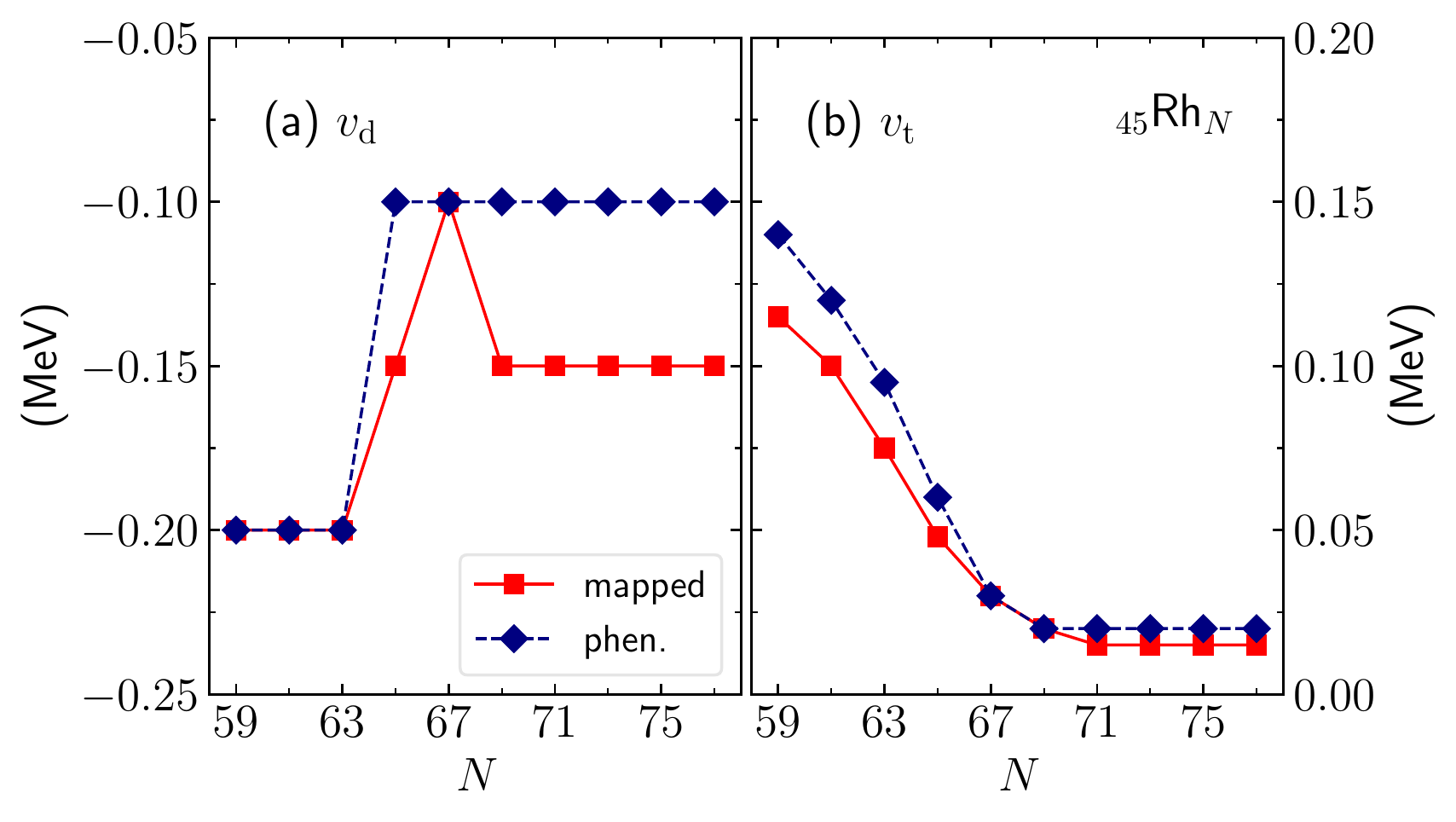}
\caption{Parameters for the residual neutron-proton 
interactions (\ref{eq:hff}) 
employed for odd-odd Rh isotopes in the mapped and phenomenological
IBM approaches.
}
\label{fig:para-doo}
\end{center}
\end{figure}


\section{Odd-odd Rh nuclei\label{sec:doo}}


The excitation energies of the low-lying positive-parity states 
obtained for odd-odd Rh isotopes are depicted in  Fig.~\ref{fig:doorh}. 
The available experimental data \cite{data} suggest that for 
$N \leqslant 71$ the ground state has spin $I^{\pi}=1^+$.
Excited $1^+$ states are also observed 
at low energy. Both the mapped [Fig.~\ref{fig:doorh}(a)] and 
phenomenological [Fig.~\ref{fig:doorh}(b)]
IBFFM-2 calculations account for the ground-state spin $1^+$. 
The calculations also reproduce reasonably well the energies
of the $1^+_2$ states. From $N \approx 71$ to 73, both 
types of calculations suggest a change in the ground-state 
spin to $I^\pi=5^+$. There are no spectroscopic data to 
compare with for even-$A$ Rh isotopes with $N \geqslant 73$. 
Note, that a ground-state spin different 
from $I^\pi=1^+$ is experimentally found 
in the neighboring odd-odd Ag and In isotopes. 
For instance, for $^{120}$Ag, $^{122}$Ag, $^{124}$Ag 
and $^{126}$In the ground state
has spin $I^{\pi}=3^+$. A low-lying 
$5^+$ level is observed in $^{122}$In at an 
excitation energy  around 40 keV above the $1^+$ ground state. 

The strength parameters $\vd$ and $\vt$ of the 
neutron-proton residual interaction $\hff$ in Eq.~(\ref{eq:hff})
are shown in Fig.~\ref{fig:para-doo} for odd-odd Rh
isotopes as functions of the neutron number. Those 
parameters are determined so that
the correct ground-state spin  $I^\pi_{g.s.}=1^+_1$
as well as the energy of the $1^+_2$ state 
are  reproduced reasonably well. 
For $N \geqslant 73$, 
where experimental data are not available, 
the same values of the parameters as for
$^{116}$Rh$_{71}$ are employed. As can be seen from
Fig.~\ref{fig:para-doo}(a), the parameter 
$\vd$ changes suddenly  from $N=63$ to 67. This 
sudden change accounts for the experimental
[see Fig.~\ref{fig:doorh}(c)]
lowering of the $1^+_2$ level toward the middle of 
the major shell, $N \approx 67$. On the other hand,
the tensor interaction strength exhibits a smooth decrease with $N$.

\begin{figure*}[ht]
\begin{center}
\includegraphics[width=.7\linewidth]{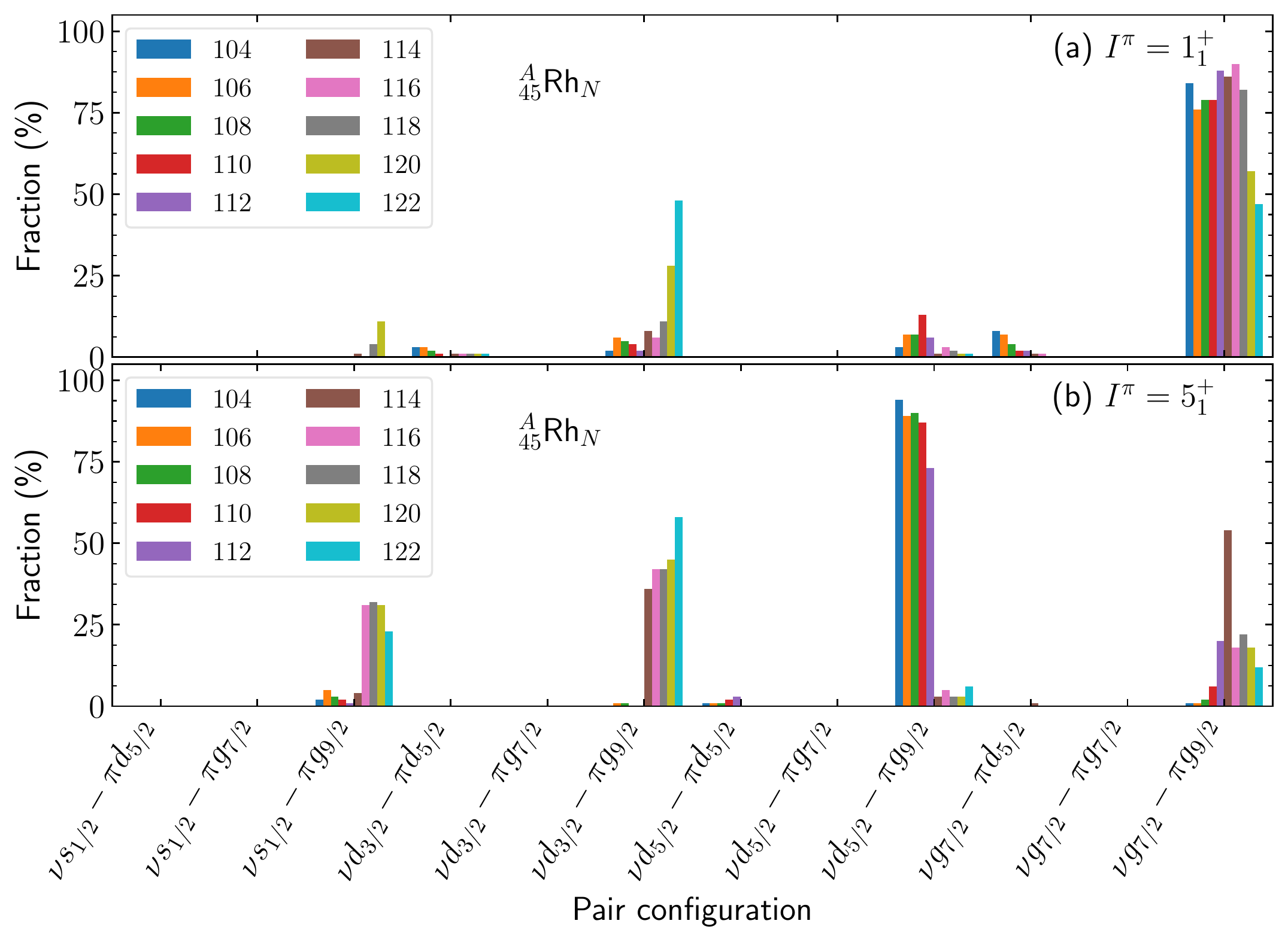}
\caption{Fraction (in percent \%) 
of the neutron-proton pair components 
in the wave functions for the 
(a) $1^+_1$ and (b) $5^+_1$ states of the odd-odd 
$^{104-122}$Rh isotopes 
under study. The wave functions are obtained within
the mapped IBFFM-2 formalism based on the 
Gogny-D1M EDF.}
\label{fig:doorh-wf}
\end{center}
\end{figure*}

The nature of the low-lying states in 
odd-odd Rh isotopes can be analyzed in terms of 
various neutron-proton 
pair components in the IBFFM-2 wave functions. 
The corresponding results for the $1^+_1$ and $5^+_1$ 
states, obtained within the mapped IBFFM-2
formalism, are shown in Fig.~\ref{fig:doorh-wf}.
For nuclei with $A \leqslant 118$, the $1^+_1$ state 
is  mostly based  on the configuration associated 
with the  $[\nu 0g_{7/2} \otimes \pi 0g_{9/2}]^{(J)}$ 
neutron-proton pairs coupled to the even-even 
boson core, with the total angular momentum of the 
fermion system $J=1,2,\ldots$, 8. 
For $^{120}$Rh and $^{122}$Rh, the contributions of the 
$[\nu 1d_{3/2} \otimes \pi 0g_{9/2}]^{(J)}$ 
($J=3,4,5,6$) pairs also play a prominent role. 
As one can see from Fig.~\ref{fig:doorh-wf}(b), 
the dominant contribution to the $5^+_1$ wave function 
for  Rh isotopes with  mass $A \leqslant 112$ 
comes from the $[\nu 1d_{5/2} \otimes \pi 0g_{9/2}]^{(J)}$ 
pair components, while 
the $[\nu 0g_{7/2} \otimes \pi 0g_{9/2}]^{(J)}$ 
pair components play a negligible role. 
For heavier Rh isotopes, with $A \geqslant 114$, 
the other pair components that involve the $\pi 0g_{9/2}$ 
state, i.e., those based on the 
$[\nu 2s_{1/2} \otimes \pi 0g_{9/2}]^{(J)}$, 
$[\nu 1d_{3/2} \otimes \pi 0g_{9/2}]^{(J)}$, and 
$[\nu 0g_{7/2} \otimes \pi 0g_{9/2}]^{(J)}$ 
pairs, are rather fragmented 
in the $I^\pi=5^+_1$ wave functions. 
Qualitatively similar results are obtained 
using  phenomenological IBFFM-2 
wave functions.

%
\begin{table}
\caption{\label{tab:doorh-em}
$B(E2)$, $B(M1)$ (in W.u.), and magnetic 
dipole moment $\mu(1^+_1)$ (in $\mu_N$) for 
odd-odd Rh isotopes, computed within the mapped IBFFM-2 
based on the Gogny-D1M EDF and the phenomenological IBFFM-2. 
Experimental data are taken from Refs.~\cite{data,stone2005}. 
}
 \begin{center}
 \begin{ruledtabular}
  \begin{tabular}{lcccc}
 & & \multicolumn{2}{c}{Calc.} & \\
\cline{3-4}
& & mapped & phen. & Expt. \\
\hline
$^{104}$Rh
& $B(E2;1^+_{3}\to2^+_{1})$ & $1.35$ & $13$ & $>5.2$ \\
& $B(M1;2^+_{1}\to1^+_{1})$ & $0.03$ & $0.06$ & $>0.029$ \\
& $B(M1;1^+_{3}\to1^+_{1})$ & $0.03$ & $0.05$ & $>0.00098$ \\
$^{106}$Rh
& $\mu(1^+_{1})$ & $2.13$ & $2.20$ & $2.575(7)$ \\
 \end{tabular}
 \end{ruledtabular}
 \end{center}
\end{table}

The experimental information on the 
electromagnetic properties of the considered odd-odd Rh
nuclei is rather limited.
Table~\ref{tab:doorh-em} compares the predicted and 
experimental $B(E2)$, $B(M1)$, and magnetic dipole 
moment $\mu(1^+_1)$ for $^{104}$Rh and $^{106}$Rh. 
Both the mapped and phenomenological 
IBFFM-2 calculations  
provide a reasonable description of the 
experimental data for these odd-odd nuclei. 
Nevertheless, a more detailed 
assessment of the quality of the IBFFM-2 wave functions 
is difficult in this case, due to the lack of data.

\begin{figure}[ht]
\begin{center}
\includegraphics[width=.9\linewidth]{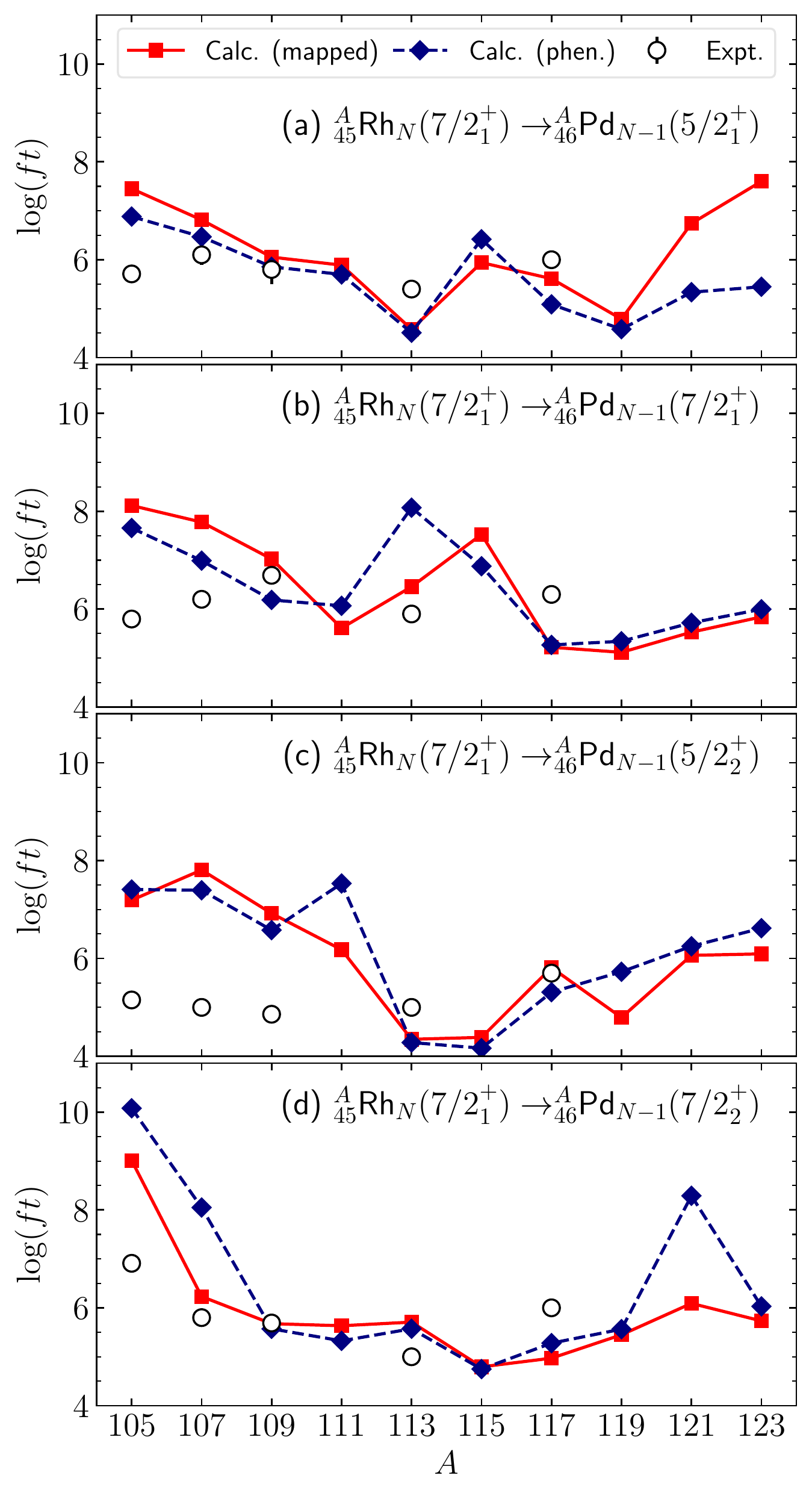}
\caption{$\ft$ values for the 
$\btm$ decays from the odd-$A$ Rh into Pd nuclei, 
(a) $7/2^+_1\to5/2^+_1$, (b) $7/2^+_1\to 7/2^+_1$, 
(c) $7/2^+_1\to5/2^+_2$, and (d) $7/2^+_1\to 7/2^+_2$ 
computed using wave functions obtained 
within the mapped and phenomenological 
IBFM-2  models. The available experimental 
data \cite{data} are also included in the plot.}
\label{fig:ft-odd}
\end{center}
\end{figure}

%
\begin{table}
\caption{\label{tab:ft-odd}
$\ft$ values for the $\btm$ 
decays from odd-$A$ Rh into Pd nuclei,
computed using wave functions obtained 
within the mapped  IBFM-2 scheme based 
on the Gogny-D1M EDF and within 
the phenomenological IBFM-2 model.
The experimental data are taken from Ref.~\cite{data}.
}
 \begin{center}
 \begin{ruledtabular}
  \begin{tabular}{lcccc}
& & \multicolumn{2}{c}{Calc.} & \\
\cline{3-4}
Decay & $I_i \to I_f$ & mapped & phen. & Expt. \\
\hline
$^{105}$Rh$\to^{105}$Pd
& ${7/2}^{+}_{1}\to{5/2}^{+}_{1}$ & 7.45 & 6.88 & 5.710(7) \\
& ${7/2}^{+}_{1}\to{7/2}^{+}_{1}$ & 8.12 & 7.66 & 5.797(16) \\
& ${7/2}^{+}_{1}\to{5/2}^{+}_{2}$ & 7.19 & 7.41 & 5.152(20) \\
& ${7/2}^{+}_{1}\to{7/2}^{+}_{2}$ & 9.01 & 10.08 & 6.91(3) \\
[1.0ex]
$^{107}$Rh$\to^{107}$Pd
& ${7/2}^{+}_{1}\to{5/2}^{+}_{1}$ & 6.81 & 6.47 & 6.1(2) \\
& ${7/2}^{+}_{1}\to{5/2}^{+}_{2}$ & 7.81 & 7.39 & 5.0(1) \\
& ${7/2}^{+}_{1}\to{7/2}^{+}_{1}$ & 7.78 & 6.99 & 6.2(1) \\
& ${7/2}^{+}_{1}\to{7/2}^{+}_{2}$ & 6.23 & 8.05 & 5.8(1) \\
& ${7/2}^{+}_{1}\to{5/2}^{+}_{3}$ & 8.00 & 7.45 & 6.1(1) \\
& ${7/2}^{+}_{1}\to{5/2}^{+}_{4}$ & 5.82 & 7.87 & 5.3(1) \\
[1.0ex]
$^{109}$Rh$\to^{109}$Pd
& ${7/2}^{+}_{1}\to{5/2}^{+}_{1}$ & 6.05 & 5.86 & 5.8(3) \\
& ${7/2}^{+}_{1}\to{7/2}^{+}_{1}$ & 7.02 & 6.19 & 6.69(12) \\
& ${7/2}^{+}_{1}\to{5/2}^{+}_{2}$ & 6.92 & 6.58 & 4.86(5) \\
& ${7/2}^{+}_{1}\to{7/2}^{+}_{2}$ & 5.68 & 5.57 & 5.69(6) \\
& ${7/2}^{+}_{1}\to{5/2}^{+}_{3}$ & 6.83 & 7.39 & 5.53(5) \\
& ${7/2}^{+}_{1}\to{9/2}^{+}_{1}$ & 7.32 & 6.84 & 7.26(19) \\
[1.0ex]
$^{113}$Rh$\to^{113}$Pd
& ${7/2}^{+}_{1}\to{5/2}^{+}_{1}$ & 4.58 & 4.51 & 5.4(1) \\
& ${7/2}^{+}_{1}\to{7/2}^{+}_{1}$ & 6.46 & 8.07 & 5.90(5) \\
& ${7/2}^{+}_{1}\to{5/2}^{+}_{2}$ & 4.35 & 4.28 & 5.00(4)\footnotemark[1] \\
& ${7/2}^{+}_{1}\to{7/2}^{+}_{2}$ & 5.71 & 5.57 & 5.00(4)\footnotemark[1] \\
\footnotetext[1]{$({5/2}^+,{7/2}^+)$ at 349 keV \cite{data}}
& ${7/2}^{+}_{1}\to{5/2}^{+}_{3}$ & 5.42 & 5.59 & 6.7(2)\footnotemark[2] \\
& ${7/2}^{+}_{1}\to{7/2}^{+}_{3}$ & 5.07 & 4.81 & 6.7(2)\footnotemark[2] \\
[1.0ex]
\footnotetext[2]{$({5/2}^+,{7/2}^+)$ at 373 keV based on the XUNDL datasets \cite{data}}
$^{117}$Rh$\to^{117}$Pd
& ${7/2}^{+}_{1}\to{5/2}^{+}_{1}$ & 5.61 & 5.09 & 6.0\footnotemark[3] \\
& ${7/2}^{+}_{1}\to{5/2}^{+}_{2}$ & 5.81 & 5.31 & 5.7\footnotemark[3] \\
& ${7/2}^{+}_{1}\to{5/2}^{+}_{3}$ & 4.27 & 5.34 & 5.8\footnotemark[3] \\
& ${7/2}^{+}_{1}\to{7/2}^{+}_{1}$ & 5.22 & 5.27 & 6.3\footnotemark[3] \\
& ${7/2}^{+}_{1}\to{5/2}^{+}_{4}$ & 7.64 & 4.56 & 6.3\footnotemark[3] \\
& ${7/2}^{+}_{1}\to{5/2}^{+}_{5}$ & 5.82 & 5.50 & 6.0\footnotemark[4] \\
& ${7/2}^{+}_{1}\to{7/2}^{+}_{2}$ & 4.97 & 5.28 & 6.0\footnotemark[4] \\
\footnotetext[3]{Uncertainties are not given with the $\ft$.}
\footnotetext[4]{$({5/2}^+,{7/2}^+)$ level at 436 keV, based on the 
XUNDL datasets \cite{data}. Uncertainties are not given.}
 \end{tabular}
 \end{ruledtabular}
 \end{center}
\end{table}


\section{$\beta$ decay\label{sec:beta}}



\subsection{$\beta$ decays between odd-$A$ nuclei\label{sec:bt-odd}}

Figure~\ref{fig:ft-odd} shows the $\ft$ 
values for the $\btm$ decays of the ${7/2}^+_1$ state 
of the odd-$A$ Rh into several low-lying states 
of the odd-$A$ Pd nuclei. Results are obtained 
using mapped and phenomenological 
IBFM-2 wave functions. In both cases, the predicted trend of the 
$\ft$ values, as  functions of the nucleon number,
 reflects 
the structural change in the parent and daughter 
odd-$A$ nuclei.
An illustrative example is a kink 
emerging 
at the mass $A \approx 113$ or 115 
in the predicted $\ft$ values for the 
${7/2}^+_1 \to {5/2}^+_1$ [Fig.~\ref{fig:ft-odd}(a)] 
and ${7/2}^+_1 \to {7/2}^+_1$ 
[Fig.~\ref{fig:ft-odd}(b)] decays. 
The mass number at which the kink emerges 
corresponds to the transitional region, where 
the ground-state spin changes,  observed in the 
odd-$A$ Pd daughter (see Fig.~\ref{fig:oddpd}).
The mass dependence of the predicted $\ft$ 
values is similar in the mapped and phenomenological
calculations, exception made of the results
from $A$=113 to 115 in the ${7/2}^+_1 \to {7/2}^+_1$ decay 
and from $A$=117 to 119 in the  ${7/2}^+_1 \to {5/2}^+_2$ decay. 

Both within 
the mapped and phenomenological schemes, the present calculations overestimate 
the observed $\ft$ values for the decays 
$^{105,107}$Rh$({7/2}^+_1) \to ^{105,107}$Pd$({5/2}^+_1)$ 
[Fig.~\ref{fig:ft-odd}(a)]. 
At both $A =$ 105 and 107, 
the ${5/2}^+_1$ final-state wave function 
has been shown to be almost purely made of the $\nu 1d_{5/2}$ 
configuration [see Fig.~\ref{fig:oddpd-wf}(c)], while the 
parent state ${7/2}^+_1$ is of almost pure 
$\pi 0g_{9/2}$ nature. 

The dominant contribution to the GT matrix element  
for the above ${7/2}^+_1 \to {5/2}^+_1$ decays 
indeed comes from the term that corresponds to the 
coupling of the $\nu 1d_{5/2}$ with $\pi 0g_{9/2}$ 
single-particle states, which is of the form
\begin{align}
\label{eq:dn-an+d5}
 [[\tilde d_\nu \times a_{\nu 1d_{5/2}}^\+]^{(7/2)} \times \tilde a_{\pi 0g_{9/2}}]^{(J=1)} \; .
\end{align}
The matrix element of this term is, however, 
rather small: 0.041 and $-0.091$ (0.069 and $-0.118$), 
for the $^{105}$Rh and $^{107}$Rh decays 
in the mapped (phenomenological) approach. 
There are many other terms similar to the one in 
Eq.~(\ref{eq:dn-an+d5}), but their 
matrix elements are small and cancel each other, 
leading to a small GT transition rate. 
The same is true for the 
$^{105,107}$Rh$({7/2}^+_1) \to ^{105,107}$Pd$({7/2}^+_1)$ 
decays [Fig.~\ref{fig:ft-odd}(b)]. 
In this case, the Fermi transition 
matrix is also  negligibly small.

The calculations underestimate 
the $\ft$ values for the 
$^{113}$Rh$({7/2}^+_1) \to ^{113}$Pd$({5/2}^+_1)$ 
decay. 
For this decay, approximately 75 \% and 25 \% of the wave 
function of the ${5/2}^+_1$ final state are comprised 
of the $\nu 1d_{5/2}$ and $\nu 0g_{7/2}$ configurations, 
respectively [see Fig.~\ref{fig:oddpd-wf}(c)]. 
Due to the large admixture of the $\nu 0g_{7/2}$ components 
into the ${5/2}^+_1$ state of $^{113}$Pd, 
the term that is proportional to 
\begin{align}
\label{eq:an+7_ap-9}
[a^\+_{0\nu g_{7/2}} \times \tilde a_{\pi 0g_{9/2}}]^{(1)}
\end{align}
makes a sizable contribution to the GT transition strength. 
The matrix element of this component, 
which amounts to $-0.788$ (0.850) in the mapped (phenomenological) 
calculation, is so large that the corresponding 
$\ft$ value is too small as compared 
with the experimental value.

As noted above, 
there are notable quantitative differences 
between the mapped and phenomenological predictions for the
$\ft$ values in the case of the  
$^{113}$Rh$({7/2}^+_1) \to ^{113}$Pd$({7/2}^+_1)$ decay.
The GT transition matrix element obtained 
in the phenomenological 
calculation is two orders of magnitude smaller 
than the one obtained within the mapped scheme. 
This difference stems from a subtle balance between 
matrix elements of different terms in the 
GT transition operator. 
The dominant contribution to the GT matrix element in the 
former calculation come from the term proportional to 
the expression in Eq.~(\ref{eq:an+7_ap-9}), 
and the one of the form
\begin{align}
s^\+_{\nu}[[\tilde d_{\nu} \times a^\+_{\nu 0g_{7/2}}]^{(7/2)} 
\times a_{\pi 0g_{9/2}}]^{(1)} \; . 
\end{align}
Their matrix elements are of 
the same order of magnitude, but have the opposite 
signs, hence cancellation occurs between these terms. 
The degree of the cancellation, however, is 
much smaller in the mapped calculation. 
The contribution of the Fermi matrix element is 
negligibly small in both the mapped and phenomenological
cases.

The $\ft$ values for the Rh decays into the 
non-yrast states, ${5/2}^+_2$ and ${7/2}^+_2$, 
of the odd-$A$ Pd are shown 
in Figs.~\ref{fig:ft-odd}(c) and \ref{fig:ft-odd}(d), 
respectively. The predicted $\ft$ values for the 
$^{A}$Rh$({7/2}^+_1) \to ^{A}$Pd$({5/2}^+_2)$ decay 
in the two sets of calculations 
are generally large, $\ft \gtrsim 7$ 
for  $A \lesssim 111$. In particular, they
overestimate 
the experimental values for the $^{105}$Rh, $^{107}$Rh, 
and $^{109}$Rh decays by a factor of two. 
The discrepancy could be attributed to the 
nature of the IBFM-2 wave functions and the 
components of the GT operator.
 The computed $\ft$ values for the 
$^{A}$Rh$({7/2}^+_1) \to ^{A}$Pd$({7/2}^+_2)$ decay 
in the mapped scheme are close to the 
experimental values, with an exception made of the 
$^{105}$Rh decay.

Table~\ref{tab:ft-odd} gives complementary results 
for the $\ft$ values of the $\btm$ decays 
$^A$Rh$({7/2}^+_1) \to ^A$Pd$(I^+_f)$, with  
final states other than those already discussed above.
The predicted  $\ft$ values are compared with the available 
experimental data \cite{data}.

Previous IBFM-2 
calculations \cite{yoshida2002} provided 
$\ft$ values for the 
$\btm$ decays ${7/2}^+_1 \to {5/2}^+_1$ and 
${7/2}^+_1 \to {7/2}^+_1$ in $^{105,107,109}$Rh
which are consistent with the experimental ones. 
However, for the same nuclei the values 
$\ft\approx$ 4 were 
obtained for the ${7/2}^+_1 \to {5/2}^+_2$ $\btm$ decay. 
Such $\ft$ values 
are systematically smaller than 
the experimental values and those obtained in this work. 
A more recent IBFM-2 calculation for the 
$^{115,117}$Rh$\to^{115,117}$Pd $\btm$ decay 
\cite{ferretti2020}
obtained a value $\ft$ = 5.90 for the 
${7/2}^+_1 \to {5/2}^+_1$ decay of $^{115}$Rh.
This $\ft$ value is close to the one obtained 
in this study. On the other hand, for the ${7/2}^+_1 \to {5/2}^+_1$
and ${7/2}^+_1 \to {7/2}^+_1$ decays of $^{117}$Rh, the
values $\ft$ = 6.78 and 6.68 were reported in \cite{ferretti2020}. 
They are 
approximately  20 \% larger than those 
obtained in the present work.

\begin{figure}[ht]
\begin{center}
\includegraphics[width=.9\linewidth]{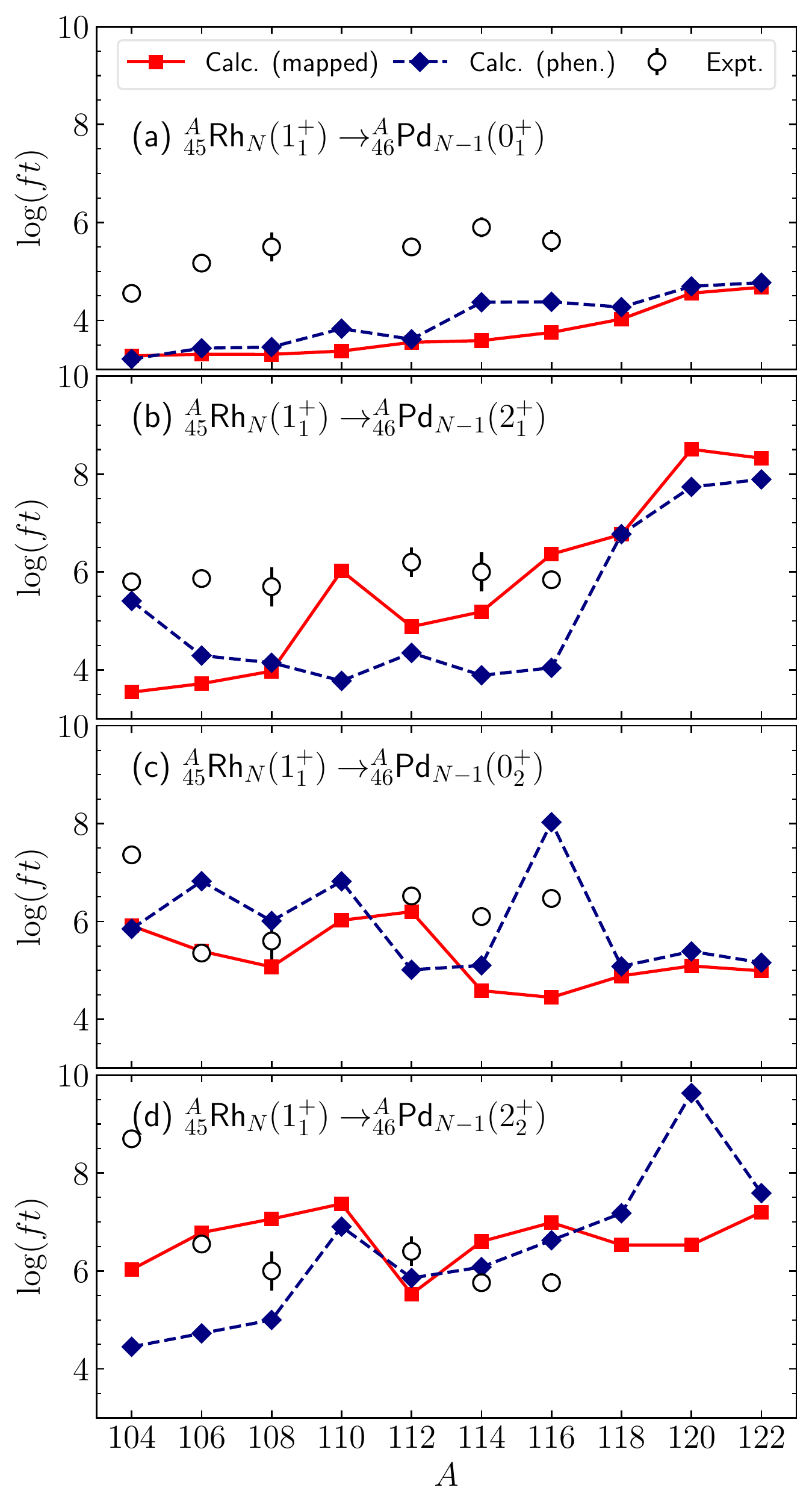}
\caption{The same as in Fig.~\ref{fig:ft-odd}, but for the 
$\btm$ decays (a) $1^+_1\to0^+_1$, (b) $1^+_1\to2^+_1$, 
(c) $1^+_1\to0^+_2$, and (d) $1^+_1\to2^+_2$ 
from the even-$A$ Rh into Pd nuclei.}
\label{fig:ft-even}
\end{center}
\end{figure}

\begin{figure}[ht]
\begin{center}
\includegraphics[width=.9\linewidth]{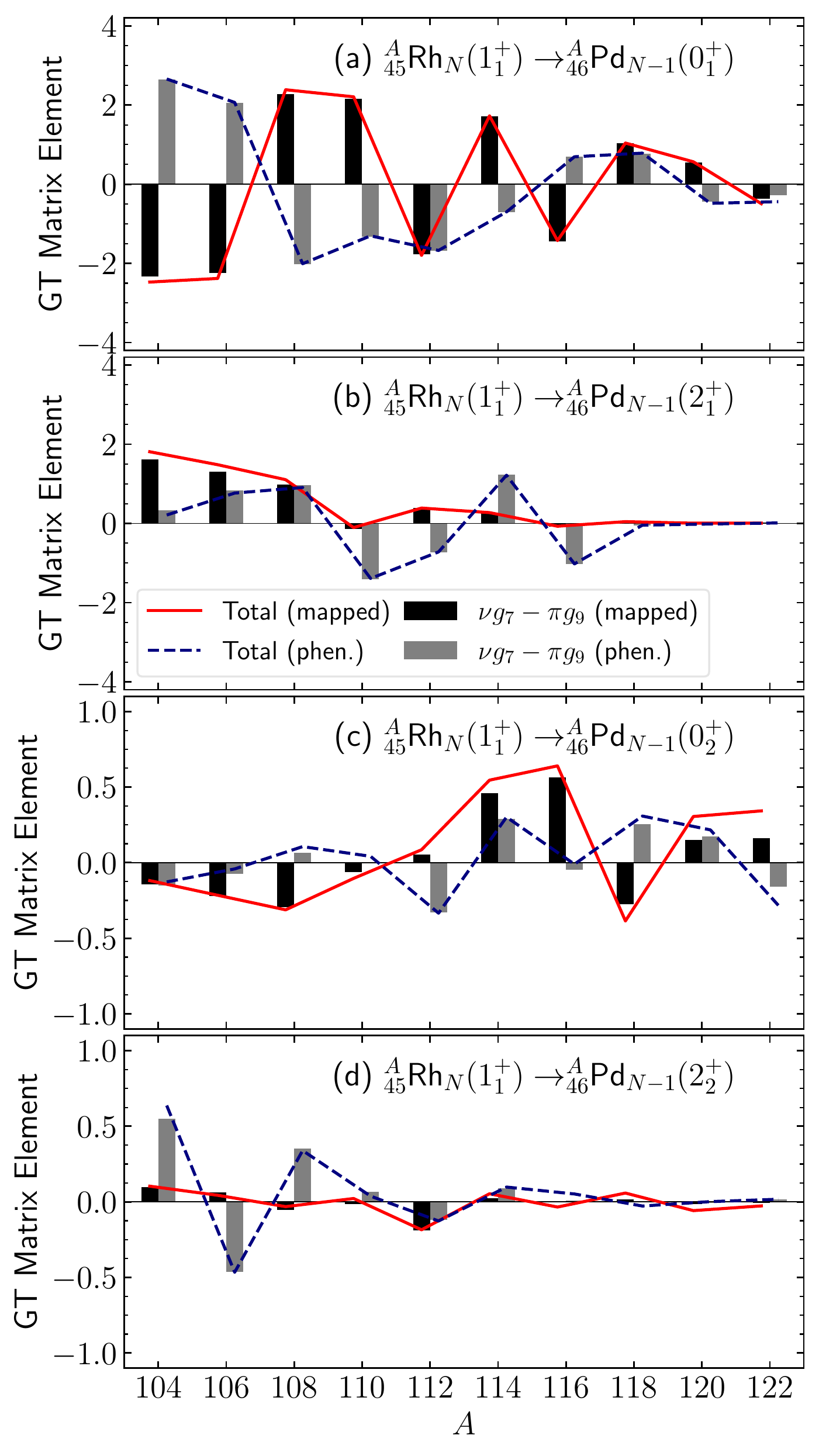}
\caption{Reduced matrix elements of the 
$\nu 0g_{7/2} - \pi 0g_{9/2}$ terms in the GT transition 
operators, and total GT matrix elements for the 
$\btm$ decays (a) ${1}^+_1 \to 0^+_1$, 
(b) ${1}^+_1 \to 2^+_1$, (c) ${1}^+_1 \to 0^+_2$, 
and (d) ${1}^+_1 \to 2^+_2$ of the even-$A$ Rh, 
resulting from the mapped and 
phenomenological calculations.}
\label{fig:ft-even-comp}
\end{center}
\end{figure}

\subsection{$\beta$ decays of even-$A$ nuclei\label{sec:bt-even}}

The $\ft$ values 
for the $\btm$ decays of the even-$A$ Rh into Pd nuclei
are plotted in Fig.~\ref{fig:ft-even}.
One immediately sees from Fig.~\ref{fig:ft-even}(a) 
that the mapped and phenomenological
$\ft$ values for the 
$^{A}$Rh$({1}^+_1) \to ^{A}$Pd$(0^+_1)$ decays 
are, approximately, a factor two  smaller 
than the experimental ones. 
The corresponding 
GT matrix elements are  almost purely determined 
by the contributions of the terms 
associated with the $\nu 0g_{7/2} - \pi 0g_{9/2}$ 
coupling, i.e.,
\begin{align}
 [\tilde a_{\nu 0g_{7/2}} \times \tilde a_{\pi 0g_{9/2}}]^{(1)} \; ,
\end{align}
for $N \leqslant 65$ and
\begin{align}
\label{eq:snap+7_ap-9}
 s_{\nu}^\+[\tilde a_{\nu 0g_{7/2}} \times \tilde a_{\pi 0g_{9/2}}]^{(1)} \; ,
\end{align}
for $N \geqslant 67$.
As shown in Fig.~\ref{fig:ft-even-comp}(a), 
the matrix elements of these terms are 
particularly large for the mass $A \leqslant 116$. 
Note also, that the IBFFM-2 wave functions for the 
initial $1^+_1$ state  mainly consist 
of the 
pair configuration 
$[\nu 0g_{7/2} \otimes \pi 0g_{9/2}]^{(J)}$ 
for the even-$A$ Rh with  $A \leqslant 118$
[see Fig.~\ref{fig:doorh-wf}(a)]. 
For the larger mass $A \geqslant 120$, this pair 
configuration becomes less important in the $1^+_1$ 
wave function of the final nucleus. 
As a consequence, the GT transition strength 
decreases with increasing $A$  
[see Fig.~\ref{fig:ft-even-comp}(a)].

To reproduce the $\beta$-decay $\ft$ data, 
effective values of the $\ga$ factor, $\gae$, 
are often employed. 
Here we compare the predicted $\ft$ value 
for the $^{A}$Rh$({1}^+_1) \to ^{A}$Pd$(0^+_1)$
decay with the corresponding 
experimental one, and extract
the $\gae$ values 
for those decays for which  $\ft$ data are available.
 The resulting $\gae$ values are, on average,  
$\gae \approx 0.152$ (0.205) in the
mapped (phenomenological) scheme. This amounts to a 
reduction of the free value by approximately 
by 88 (84) \%. 
In the previous IBM-2/IBFFM-2 study of the 
$\beta$ and $\db$ decays of the Te and Xe isotopes 
with  $A \approx 130$ \cite{yoshida2013}, 
the $\gae$ values extracted 
from a comparison with the $\ft$ data for the 
single-$\beta$ decays are 0.313 
for the $\btp$
decay $^{128}$I$(1^+_1) \to ^{128}$Te$(0^+_1)$,  
and 0.255 for the $\btm$ decay 
$^{128}$I$(1^+_1) \to ^{128}$Xe$(0^+_1)$.

As can be seen from Fig.~\ref{fig:ft-even}(b), 
the $\ft$ values obtained within the 
mapped and phenomenological approaches
for the $^{A}$Rh$({1}^+_1) \to ^{A}$Pd$(2^+_1)$ decay 
differ considerably.
The difference between the two calculations is 
especially large at $A=110$ and 116. 
One sees from Fig.~\ref{fig:ft-even-comp}(b), that
the GT matrix element 
$M(\text{GT}; {1}^+_1 \to 2^+_1)$ for the $^{116}$Rh decay 
in the phenomenological calculations
is much larger in magnitude than the one 
obtained within the mapped approach, with the largest
 contribution coming from the term 
associated with the $\nu 0g_{7/2} - \pi 0g_{9/2}$ 
coupling. Generally, the predicted $\ft$ values for the 
$1^+_1 \to 2^+_1$ $\btm$ decay, both within the 
mapped and phenomenological schemes, increase with $A$ (or $N$). 
This is due to the fact that the pair configuration 
$[\nu 0g_{7/2} \otimes \pi 0g_{9/2}]^{(J)}$ 
gradually becomes less important in the 
$1^+_1$ wave function of the even-$A$ Rh for 
larger $A$ [see Fig.~\ref{fig:doorh-wf}(a)].

For the  $^{A}$Rh$({1}^+_1) \to ^{A}$Pd$(0^+_2)$ decay, 
the $\ft$ values
predicted within the mapped and phenomenological
approaches are similar.
The most notable difference occurs at $A=116$, with
the mapped $\ft$ value being nearly half the
phenomenological one. This is a consequence of 
the fact that in the mapped GT matrix element 
$M(\text{GT}; {1}^+_1 \to 0^+_2)$ associated with
the $^{116}$Rh decay, the component of Eq.~(\ref{eq:snap+7_ap-9})
is an order of magnitude  larger 
than the one in the phenomenological calculations 
[see Fig.~\ref{fig:ft-even-comp}(c)].
In addition, the computed $\ft$ values for the 
$1^+_1 \to 0^+_2$ decay are 
larger than those for the 
$1^+_1 \to 0^+_1$ decay because
the matrix elements of the 
components involving 
the coupling $\nu 0g_{7/2} - \pi 0g_{9/2}$ in the 
$M(\text{GT}; {1}^+_1 \to 0^+_2)$ strength 
are smaller in magnitude than those 
in the $M(\text{GT}; {1}^+_1 \to 0^+_1)$ one.
 
The $\ft$ values corresponding to the  
$^{A}$Rh$({1}^+_1) \to ^{A}$Pd$(2^+_2)$ decay are 
depicted in Fig.~\ref{fig:ft-even}(d). Both the
mapped and phenomenological calculations 
largely underestimate the 
measured value at $A=104$. However, the results 
obtained with both 
schemes reproduce the experimental trend reasonably well
for $108 \leqslant A \leqslant 116$. 
As can be seen from Fig.~\ref{fig:ft-even-comp}(d), the 
difference between the mapped and 
phenomenological results for 
$104 \leqslant A \leqslant 108$  is due 
to the difference between the  matrix elements for 
the components $\nu 0g_{7/2} - \pi 0g_{9/2}$ in both 
schemes, with  the mapped matrix elements being 
an order of magnitude smaller than the phenomenological
ones.

%
\begin{table}
\caption{\label{tab:ft-even}
The same as in Table~\ref{tab:ft-odd}, but for the $\btm$ decays 
from even-$A$ Rh to Pd nuclei. 
}
 \begin{center}
 \begin{ruledtabular}
  \begin{tabular}{lcccc}
& & \multicolumn{2}{c}{Calc.} & \\
\cline{3-4}
Decay & $I_i \to I_f$ & mapped & phen. & Expt. \\
\hline
$^{104}$Rh$\to^{104}$Pd
& $1^{+}_{1}\to0^{+}_{1}$ & 3.27 & 3.21 & 4.55(1) \\
& $1^{+}_{1}\to2^{+}_{1}$ & 3.54 & 5.41 & 5.80(1) \\
& $1^{+}_{1}\to0^{+}_{2}$ & 5.91 & 5.85 & 7.36(2) \\
& $1^{+}_{1}\to2^{+}_{2}$ & 6.03 & 4.45 & 8.7(1) \\
& $1^{+}_{1}\to0^{+}_{3}$ & 6.42 & 6.05 & 5.5(1) \\
& $1^{+}_{1}\to2^{+}_{3}$ & 5.24 & 4.72 & 6.3(1) \\
& $5^{+}_{1}\to4^{+}_{1}$ & 7.26 & 8.30 & 7.3(1) \\
& $5^{+}_{1}\to4^{+}_{2}$ & 8.45 & 7.59 & 6.1(1) \\
& $5^{+}_{1}\to4^{+}_{3}$ & 8.06 & 8.04 & 6.2(1) \\
& $5^{+}_{1}\to4^{+}_{4}$ & 8.59 & 8.57 & 5.8(1) \\
[1.0ex]
$^{106}$Rh$\to^{106}$Pd
& $1^{+}_{1}\to0^{+}_{1}$ & 3.31 & 3.43 & 5.168(7) \\
& $1^{+}_{1}\to2^{+}_{1}$ & 3.72 & 4.29 & 5.865(17) \\
& $1^{+}_{1}\to2^{+}_{2}$ & 6.78 & 4.72 & 6.55(7) \\
& $1^{+}_{1}\to0^{+}_{2}$ & 5.39 & 6.82 & 5.354(19) \\
& $1^{+}_{1}\to2^{+}_{3}$ & 5.15 & 4.58 & 5.757(17) \\
[1.0ex]
$^{108}$Rh$\to^{108}$Pd
& $1^{+}_{1}\to0^{+}_{1}$ & 3.31 & 3.45 & 5.5(3) \\
& $1^{+}_{1}\to2^{+}_{1}$ & 3.97 & 4.14 & 5.7(4) \\
& $1^{+}_{1}\to2^{+}_{2}$ & 7.06 & 5.00 & 6.0(4) \\
& $1^{+}_{1}\to0^{+}_{2}$ & 5.07 & 6.01 & 5.6(4) \\
& $5^{+}_{1}\to6^{+}_{1}$ & 7.72 & 7.44 & 6.8(3) \\
& $5^{+}_{1}\to4^{+}_{2}$ & 8.28 & 7.00 & 4.84(9)\footnotemark[1] \\
& $5^{+}_{1}\to5^{+}_{1}$ & 9.59 & 8.35 & 4.84(9)\footnotemark[1] \\
& $5^{+}_{1}\to6^{+}_{2}$ & 9.30 & 9.42 & 4.84(9)\footnotemark[1] \\
[1.0ex]
\footnotetext[1]{${4}^+,{5}^+,6^+$ level at 2864 keV}
$^{110}$Rh$\to^{110}$Pd
& $6^{+}_{1}\to6^{+}_{1}$ & 8.29 & 8.26 & 6.38(13) \\
& $6^{+}_{1}\to6^{+}_{2}$ & 9.57 & 8.95 & 7.1(4) \\
& $6^{+}_{1}\to5^{+}_{1}$ & 9.16 & 8.69 & 6.34(25) \\
[1.0ex]
$^{112}$Rh$\to^{112}$Pd
& $1^{+}_{1}\to0^{+}_{1}$ & 3.55 & 3.61 & $\approx$5.5 \\
& $1^{+}_{1}\to2^{+}_{1}$ & 4.88 & 4.35 & 6.2(3) \\
& $1^{+}_{1}\to2^{+}_{2}$ & 5.53 & 5.86 & 6.4(3) \\
& $1^{+}_{1}\to0^{+}_{2}$ & 6.20 & 5.01 & 6.52(6) \\
& $1^{+}_{1}\to0^{+}_{3}$ & 7.48 & 6.36 & 6.88(9)\footnotemark[2] \\
& $1^{+}_{1}\to1^{+}_{1}$ & 7.74 & 5.66 & 6.88(9)\footnotemark[2] \\
& $1^{+}_{1}\to2^{+}_{3}$ & 5.83 & 5.39 & 6.88(9)\footnotemark[2] \\
\footnotetext[2]{$(0,1,2)^+$ level at 1140 keV}
& $1^{+}_{1}\to2^{+}_{3}$ & 5.83 & 5.39 & 6.97(22) \\
& $1^{+}_{1}\to2^{+}_{3}$ & 5.83 & 5.39 & 6.50(7) \\
& $6^{+}_{1}\to6^{+}_{1}$ & 8.75 & 8.80 & 6.52\footnotemark[3] \\
& $6^{+}_{1}\to5^{+}_{1}$ & 8.96 & 10.34 & 6.54 \\
& $6^{+}_{1}\to6^{+}_{2}$ & 9.15 & 8.82 & 6.88 \\
[1.0ex]
\footnotetext[3]{$\ft$ values should be considered approximate \cite{data}.}
$^{114}$Rh$\to^{114}$Pd
& $1^{+}_{1}\to0^{+}_{1}$ & 3.59 & 4.37 & 5.9(2) \\
& $1^{+}_{1}\to2^{+}_{1}$ & 5.19 & 3.89 & 6.0(4) \\
& $1^{+}_{1}\to2^{+}_{2}$ & 6.60 & 6.08 & 5.7(2) \\
& $1^{+}_{1}\to0^{+}_{2}$ & 4.59 & 5.10 & 6.1(2) \\
& $1^{+}_{1}\to2^{+}_{3}$ & 5.57 & 5.28 & 6.1(2) \\
[1.0ex]
$^{116}$Rh$\to^{116}$Pd
& $1^{+}_{1}\to0^{+}_{1}$ & 3.75 & 4.38 & 5.62(22) \\
& $1^{+}_{1}\to2^{+}_{1}$ & 6.36 & 4.04 & 5.84(18) \\
& $1^{+}_{1}\to2^{+}_{2}$ & 6.99 & 6.63 & 5.76(19) \\
& $1^{+}_{1}\to0^{+}_{2}$ & 4.45 & 8.03 & 6.47(20) \\
& $1^{+}_{1}\to0^{+}_{3}$ & 5.29 & 8.60 & 6.36(19) \\
& $1^{+}_{1}\to2^{+}_{3}$ & 5.05 & 5.00 & 6.81(21) \\
  \end{tabular}
 \end{ruledtabular}
 \end{center}
\end{table}

For the sake of completeness, 
Table~\ref{tab:ft-even} compares the predicted 
and experimental \cite{data} $\ft$ values 
for the $\btm$ decays 
of the even-$A$ Rh isotopes. Cases other than 
hose already discussed above are considered 
in the table. 
As compared with the ground-state-to-ground-state
decay $1^+_1 \to 0^+_1$, the $ft$ values for 
the decays of the $1^+_1$ state into non-yrast 
$1^+$ and $2^+$ states, 
and the $\ft$ values for the $5^+_1 \to I_f$ and $6^+_1 \to I_f$ 
decays  are calculated to be large.
Note that the predicted $\ft$ values for the decays 
$^{104}$Rh$(5^+_1) \to ^{104}$Pd$(4^+_1)$ 
and 
$^{108}$Rh$(5^+_1) \to ^{108}$Pd$(6^+_1)$ 
are rather close to the experimental ones. 


\section{Conclusions\label{sec:summary}}


In this paper, the low-energy collective states 
and $\beta$ decays  for 
even and odd-mass neutron-rich Rh and Pd isotopes
have been studied using a mapping  framework based on the 
Gogny-EDF and the particle-boson coupling scheme.
The constrained HFB has been employed to provide 
microscopic input to the mapping procedure. Such
an input consists of potential energy surfaces 
as functions of the  $(\beta,\gamma)$ shape degrees of 
freedom for the  even-even $^{104-124}$Pd isotopes.
The  IBM-2 Hamiltonian, used to describe 
even-even
core nuclei, has been determined 
by mapping the Gogny-D1M HFB fermionic potential
energy surfaces onto the corresponding bosonic surfaces.
The microscopic mean-field calculations also 
provided single-particle energies for the odd systems. 
Those represent essential building blocks of the boson-fermion 
interactions for the neighboring odd-$A$ and 
odd-odd nuclei as well as for  the 
GT and Fermi transition operators. 
The strength parameters of the 
boson-fermion  and residual 
neutron-proton interactions were fitted 
to low-energy data for the odd-$A$ and odd-odd systems.

The Gogny-HFB $(\beta,\gamma)$ potential energy 
surfaces obtained for even-even Pd isotopes point 
towards a  transition from prolate deformed 
($^{104-108}$Pd) 
to $\gamma$-soft ($^{110-116}$Pd), and to 
nearly spherical shapes ($^{118-124}$Pd). 
The low-energy excitation spectra 
and $B(E2)$ transition 
strengths resulting from the diagonalization of the 
mapped IBM-2 Hamiltonian 
reproduced the experimental trends reasonably well
and reflect, to a large extent, 
the structural evolution of the ground-state shapes 
predicted at the mean-field level.
The excitation energies obtained for the 
low-lying positive-parity levels in the 
odd-$A$ Pd and Rh, and even-$A$ Rh nuclei also 
exhibit signatures of this structural evolution. Within
this context, a notable example is the 
change in the ground state spin from 
$^{113}$Pd to $^{115}$Pd.
The computed $\ft$ values for the $\btm$ decays 
of the odd- and even-$A$ Rh into Pd nuclei have been shown 
to be sensitive to the nature of the wave functions 
of the parent and daughter nuclei. They also reflect 
the rapid structural evolution along the considered 
isotopic chains. 
The $\ft$ values for the 
odd-$A$ Rh decay have been predicted to be  
larger than the experimental ones for 
$A \lesssim 109$. This could be traced back to 
the structure of the IBFM-2 wave functions for the 
odd-$A$ daughter (Pd) nuclei. 
Furthermore, it has been shown that 
for the even-$A$ Rh decay, the 
neutron-proton pair components 
$[\nu 0g_{7/2} \otimes \pi 0g_{9/2}]^{(J)}$ 
play a key role in  the 
GT transition matrix elements and are responsible 
for the too small $\ft$ values for 
the $^A$Rh$(1^+_1) \to ^{A}$Pd$(0^+_1)$ decay 
with respect to the experimental data.

The results of the mapped calculations 
have been compared with conventional 
IBM-2 calculations in which the parameters for the 
boson Hamiltonian have been fit to the experiment.
The mapped and phenomenological  IBM-2
excitation spectra for
even-even, odd-$A$, and odd-odd systems are 
similar. However, the two sets of calculations
differ in their predictions for
electromagnetic 
and $\beta$-decay properties of the odd-nucleon systems. 
 
The results obtained in this study could be 
considered a plausible step towards a 
consistent simultaneous  
description of the  low-lying states and 
$\beta$-decay properties of atomic nuclei.
However, the difference 
between the predicted 
and experimental $\beta$-decay $\ft$ values 
might require additional 
refinements of the employed theoretical framework.
 In particular, the small $\ft$ values obtained suggest that
 the role of the effective axial-vector 
coupling constant $\ga$ should be further studied 
in future calculations. The $\gae$ values extracted in
this work from the comparison with the experimental data 
turned 
out to be by a factor 7-8 smaller than the free 
nucleon value.  This large quenching  
indicates deficiencies in the model space 
of the calculations or of the theoretical 
procedure itself. Investigation along these 
lines is in progress and will be reported 
 elsewhere.

\acknowledgments
This work is financed within the Tenure Track Pilot Programme of 
the Croatian Science Foundation and the \'Ecole Polytechnique 
F\'ed\'erale de Lausanne, and Project No. TTP-2018-07-3554 Exotic 
Nuclear Structure and Dynamics, with funds of the Croatian-Swiss 
Research Programme.
The work of LMR is supported by the Spanish 
Ministry of Economy and Competitiveness (MINECO) Grant 
No. PGC2018-094583-B-I00.

\bibliography{refs}

\end{document}